\documentclass[11pt]{article}

\usepackage[margin=1in]{geometry}
\usepackage{graphicx}
\usepackage{subcaption}
\usepackage{epstopdf,epsfig}
\usepackage{newtxtext}
\usepackage{newtxmath}
\usepackage{natbib}
\usepackage{amsmath}
\usepackage{bm}
\usepackage{siunitx}
\usepackage{booktabs}
\usepackage{mathtools}
\usepackage{hyperref}

\hypersetup{
    colorlinks = true,
    urlcolor   = blue,
    citecolor  = black,
    linkcolor  = black
}

\newcommand{\appref}[1]{\hyperref[#1]{Appendix~\ref*{#1}}}

\title{Rarefaction-induced inflation and similarity breakdown of hypersonic bow shocks over a circular cylinder}

\author{%
Ehsan Roohi\thanks{Corresponding author: \href{mailto:roohie@umass.edu}{roohie@umass.edu}}\\
\small Department of Mechanical and Industrial Engineering, University of Massachusetts Amherst,\\
\small 160 Governors Dr., Amherst, MA 01003, USA
\and
Ahmad Shoja-Sani\\
\small Department of Mechanical Engineering, Ferdowsi University of Mashhad,\\
\small Mashhad 9177948974, Iran
}

\date{}

\begin{document}

\maketitle

\begin{abstract}
Rarefied hypersonic bow shocks over blunt bodies inflate as the Knudsen number
increases, but it remains unclear whether this inflation is a simple shift and
broadening of one common shock layer or a multi-scale change of the macroscopic
and internal-energy fields. We address this question using direct simulation
Monte Carlo (DSMC) data for Mach-10 flow over a circular cylinder in argon and
nitrogen over \(Kn_\infty \approx 0.01\)--\(1\), together with a Mach-number
sweep at \(Kn_\infty=0.01\). At low rarefaction, a ray-based density-gradient
ridge gives a reproducible bow-shock location and agrees with an independent
schlieren-based shock-wave-detection method. As \(Kn_\infty\) increases, this
ridge is replaced by a broad kinetic compression layer, so the high-Knudsen
cases are analysed using profile-based standoff and thickness metrics rather
than by imposing a visual shock line. The Knudsen- and Mach-number sweeps
separate two mechanisms. At fixed \(M_\infty\), the continuum normal-shock
density ratio provides a useful low-rarefaction reference compression scale,
whereas the measured standoff growth is governed primarily by the kinetic mean
free path; the effective density thickness shows an intermediate minimum before
increasing in the diffuse regime. At fixed low \(Kn_\infty\), changing \(M_\infty\) mainly
changes compression strength and curvature, preserving a coherent attached-layer
structure. Density-registered profiles and shock-attached proper orthogonal
decomposition (POD) show that, within the present maximum-density-gradient
registration, density becomes nearly rank one, whereas Mach number and thermal
variables retain independent modal content. Rarefied
bow-shock inflation is therefore a coupled compression--relaxation process, not
a single-scale rescaling of a continuum-like shock.
\end{abstract}

\section{Introduction}
\label{sec:introduction}

Hypersonic flow over blunt bodies is governed by the coupled action of shock
compression, boundary-layer development, surface accommodation and
aerothermodynamic loading. When the molecular mean free path becomes comparable
with the body length scale, the continuum approximation progressively loses
validity and the bow shock can no longer be interpreted as a mathematical
discontinuity. Instead, the compression region has a finite kinetic thickness
whose location, width and internal relaxation structure depend on
\(Kn_\infty\), gas model and surface interaction. This rarefied regime is
central to high-altitude flight, planetary entry, plume--body interaction and
the validation of kinetic solvers. Direct simulation Monte Carlo (DSMC) remains
the reference particle method for such flows
\citep{Bird1994,Cercignani2000,Roohi2025DSMCBook}, while kinetic-theory and
asymptotic analyses provide the complementary framework for interpreting
slip, jump and high-Knudsen-number transport near solid bodies
\citep{Sone2007,Karniadakis2005,SharipovKalempa2003,Sharipov2011}.

Circular cylinders and spheres have long served as canonical blunt-body
geometries in rarefied-gas dynamics because they contain a stagnation-line
compression layer, a curved detached bow shock and strong gas--surface
nonequilibrium in a geometrically simple configuration. For stationary
cylinders, DSMC calculations by \citet{Stefanov2000Fluctuations} used the
hypersonic circular-cylinder problem to study macroscopic fluctuations in
rarefied flow fields, emphasizing that particle noise and statistical
uncertainty are not merely numerical details but part of the diagnostic
difficulty in extracting shock-layer quantities. The present test case is  motivated by the continuum-breakdown study of
\citet{Lofthouse2007ContinuumBreakdown}, who compared computational fluid dynamics (CFD) and DSMC predictions
for Mach-10 argon flow over a 12-in. circular cylinder and showed that
gradient-length local Knudsen-number regions in the shock, boundary layer and
wake can significantly affect aerothermodynamic quantities as the flow becomes
more rarefied. In the slip and early
transition regimes, \citet{Lofthouse2008SlipJump} showed that velocity slip, temperature jump and
non-equilibrium effects can strongly alter surface heating, force components
and wake structure. 

Hybrid particle--continuum simulations by
\citet{Schwartzentruber2007Modular} and related high-fidelity DSMC studies
further demonstrate that rarefied hypersonic aerothermodynamics is governed by
localized nonequilibrium and gas--surface interaction effects rather than by
continuum shock relations alone.  Related DSMC cylinder
benchmarks were later used to validate new collision algorithms and
hypersonic nonequilibrium simulations
\citep{Goshayeshi2015SBTTAS,Goshayeshi2015SBT}. More recent multiscale kinetic
methods, including unified gas-kinetic scheme (UGKS) and unified gas-kinetic wave-particle method (UGKWP), have also used hypersonic cylinder cases as
tests for all-regime rarefied-flow algorithms, including Mach-20 and Mach-30
cylinder problems spanning continuum and free-molecular limits
\citep{Huang2012UGKS,Zhu2019UGKWP,Liu2019UGKWP}. These studies establish the
cylinder as a standard benchmark for force, heat-transfer and solver-validation
questions, but they do not directly answer how the detached bow-shock layer
itself loses single-scale similarity as rarefaction increases. Recent DSMC studies of wedge--cylinder
shock--shock interaction over a wide Knudsen-number range similarly show that
increasing rarefaction weakens the ability of the wave system to deflect and
concentrate energy, altering heat-flux and skin-friction augmentation over the
cylinder surface \citep{Jiang2024ShockShock}. These works motivate the present
analysis, but their primary observables are force, heating, slip or interaction
patterns, not the modal structure of the inflated bow-shock layer.

A second body of work has examined cylinders with rotation or moving surfaces,
where rarefaction modifies the classical Magnus-effect picture and changes the
balance of lift, drag and wall shear. \citet{Riabov1999SpinningCylinder}
investigated the aerodynamics of spinning cylinders in rarefied gas flows, and
subsequent studies of high-speed rarefied flow past rotating cylinders reported
inverse-Magnus behaviour and nonequilibrium force decomposition
\citep{John2016InverseMagnus,John2018SpinningCylinder,John2019Cylinder}. 

Shock detection is itself a separate issue in rarefied hypersonic flows. In
continuum-like regimes, a schlieren or density-gradient edge gives a useful
visual shock location. At larger \(Kn_\infty\), however, the shock becomes a
finite-thickness kinetic compression layer whose apparent centre depends on the
diagnostic variable and on the extraction method. The shock-wave-detection (SWD)
technique of \citet{Akhlaghi2017SWD} provided a systematic way to identify
shock centres and finite-thickness shock sides from numerical schlieren fields,
and \citet{Akhlaghi2021ShockPolar} applied this idea to shock-polar analysis in
rarefied flow over a circular cylinder. These methods are valuable for
validating a shock-front extraction when a distinct edge exists, but they also
highlight the limitation of representing transitional and free-molecular
bow-shock layers by a single curve.

The present use of shock location and thickness measures should be interpreted
in this operational sense. We do not seek a universal definition of a true shock
surface or a universal shock thickness in the transitional and free-molecular
regimes. Instead, the density-gradient location, gradient thickness and
attached-coordinate maps are diagnostic quantities used to compare different
macroscopic and internal-energy fields under the same registration procedure.
The question is therefore not whether finite shock thickness or rarefied
relaxation exists in general, but whether a curved, detached hypersonic
compression layer admits one common attached similarity coordinate for density,
momentum and thermal variables.

This framing separates the present question from three neighbouring bodies of
work. Classical rarefied-gas analyses and slip/jump studies, including the
Boltzmann-based framework of \citet{Cercignani2000}, the molecular-gas-dynamics
treatment of \citet{Sone2007}, and the gas--surface slip and jump data of
\citet{SharipovKalempa2003,Sharipov2011}, provide the asymptotic framework for
near-wall rarefied layers and boundary conditions, but they do not by themselves
decide whether a detached, curved, external compression layer admits a single
attached coordinate common to density and thermal variables. Hornung-type detached-shock and standoff scalings provide
the appropriate continuum and near-continuum reference for the Mach sweep and
for the low-
\(Kn_\infty\) limit, but they do not determine the variable-dependent
post-shock relaxation structure inside a finite rarefied layer. Existing DSMC
cylinder benchmarks establish forces, heating, continuum breakdown and solver
validation; the present analysis uses the same canonical geometry to ask a
different question about selective compactness and similarity breakdown of the
registered shock-layer fields.

Recent data-driven studies provide another motivation for revisiting the
structure of rarefied cylinder flows. Neural-network and neural-operator
surrogates can represent DSMC fields over Mach- and Knudsen-number sweeps, but
their success depends on whether the underlying flow family has a compact
low-dimensional structure \citep{Roohi2026PoF,Roohi2026NeuralCollisionOperators}. In a related rarefied
micro-nozzle problem, \citet{RoohiMahdavi2026Nozzle} showed that a nearly
one-dimensional internal compression layer becomes highly compact after
shock-centred registration: the leading density proper orthogonal decomposition (POD) mode increased from
\(83.33\%\) of the fluctuation energy in physical coordinates to \(98.33\%\)
in a jump-scaled coordinate, and a two-dimensional shock-window POD retained
\(99.05\%\) of the energy in the first two modes. That result demonstrates
that for a nominally one-dimensional compression layer, most of the apparent
parametric complexity can be removed by locating the density-gradient station
and scaling by the finite shock thickness. The present cylinder problem asks a
more demanding question: whether the same shock-centred compactness survives
for a curved, detached bow-shock layer with angular variation, wall curvature
and gas-dependent thermal relaxation. Modal-analysis methods such as POD provide a direct way to measure this
compactness: snapshots are arranged in a data matrix and decomposed by
singular-value analysis, so that the cumulative energy quantifies the number of
coherent structures needed to represent the parameter-induced variation
\citep{Lumley1967,Sirovich1987,Berkooz1993,Taira2017Modal}. In contrast,
dynamic mode decomposition targets time-shifted data and is used to identify
growth rates and frequencies \citep{Schmid2010,Rowley2009,Towne2018}; the
present work instead uses POD as a parameter-space compactness diagnostic.

Modal analysis has also begun to enter kinetic and DSMC descriptions of
hypersonic nonequilibrium flows. For example,
\citet{Klothakis2021KineticFlatPlate} combined DSMC base flows with linear
stability analysis for hypersonic flat-plate boundary layers, while
\citet{Senkardesler2025MolecularSecondMode} used DSMC snapshots and dynamic
mode decomposition to identify second-mode wave packets in a Mach-6 flat-plate
boundary layer. In a shock-dominated configuration,
\citet{Sawant2021KineticDoubleWedge} showed that rarefied hypersonic
shock/boundary-layer interaction can sustain coherent global-instability
structures not only in the separated region but also within the detached-shock
layer. More recently, \citet{Karpuzcu2025RampSPOD} applied spectral proper
orthogonal decomposition (SPOD) and dynamic mode decomposition (DMD) to
DSMC-generated unsteady separated ramp flows and identified coherent
shock-layer and shear-layer structures from stochastic kinetic snapshots.
These studies motivate the use of modal tools for kinetic hypersonic data. In
contrast to time-resolved SPOD/DMD analyses of unsteady disturbance fields, the
present work uses POD as a parameter-space compactness diagnostic for
shock-attached DSMC fields across Mach and Knudsen-number sweeps.

The unresolved question addressed here is therefore not whether rarefied
hypersonic cylinder flows have been simulated before; they have. The question
is whether the strong shock-centred collapse observed previously in an internal
rarefied nozzle compression layer \citep{RoohiMahdavi2026Nozzle} extends to a
geometrically more complex external bow-shock layer, or whether density,
pressure, Mach number, translational temperature and rotational temperature
retain distinct shock-layer scales after registration. This distinction is important because
many reduced-order descriptions and surrogate models implicitly assume that the
dominant effect of rarefaction is a shift or broadening of a common layer. If
density collapses but thermal and rotational fields remain multi-modal, then
the inflated bow shock is not a simple geometrically rescaled shock but a
multi-scale compression--relaxation structure.

The novelty of the present work is threefold. First, we validate a ray-based
density-gradient shock extraction against the Akhlaghi et al. SWD method only
in the low-\(Kn_\infty\) regime where a distinct shock edge exists, and we do
not impose a visual shock line in diffuse high-\(Kn_\infty\) cases. Second, we
extract quantitative profile-based standoff and thickness metrics from
body-normal rays for both Knudsen- and Mach-number sweeps, so that high-Kn cases
are treated as compression layers rather than discontinuities. Third, we use
profile-level registration gain and shock-attached POD to show that, under
the maximum-density-gradient registration used here, density compression becomes
nearly rank one, whereas Mach number and thermal variables retain additional
coherent structure. Thus the work connects
classical rarefied cylinder aerothermodynamics, shock detection and modal
compactness into a single framework for measuring the breakdown of single-scale
bow-shock similarity.

This question is tested through three explicit hypotheses. First, if
rarefaction-induced bow-shock inflation were mainly a geometric shift and
broadening, density registration should remove most of the density variance
(\(H_1\)). Second, if the density-compression scale were universal, Mach number,
pressure and the thermal/internal-energy fields should collapse with the same
coordinate (\(H_2\)); persistent modal content in those variables would instead
indicate independent relaxation scales. Third, the Knudsen-number and
Mach-number sweeps should have different compactness signatures if rarefaction
changes the kinetic nature of the layer rather than only the inviscid
compression strength (\(H_3\)).

The contribution is not a new DSMC validation of hypersonic cylinder flow,
nor a surrogate-modelling study of the type considered in recent neural-network
work on rarefied cylinder data. Rather, this canonical configuration is used to
test a specific fluid-mechanical hypothesis: whether rarefaction-induced
bow-shock inflation can be represented by a single shock-attached coordinate
common to density, momentum and thermal variables. The answer is selectively
negative. The density field admits an almost rank-one registered
representation, pressure follows the density coordinate more closely than the
thermal variables, whereas Mach number and translational/rotational
temperatures retain additional coherent modal content. This selective collapse
is the central physical result: one-coordinate descriptions are useful for the
density-compression layer but are not generally sufficient for nonequilibrium
thermal/internal-energy fields. Rarefied bow-shock inflation is therefore a
coupled compression--relaxation process rather than a geometric rescaling of a
continuum-like shock layer.

The paper is organized as follows. Section~\ref{sec:problem} defines the DSMC
data sets and the shock-attached scaling. Section~\ref{sec:results_discussion}
presents the results and discussion, including rarefaction-induced density
inflation, Mach-number dependence, quantitative shock-layer metrics and POD
compactness. Section~\ref{sec:conclusions} summarizes the main conclusions.

\section{Problem definition and DSMC data}
\label{sec:problem}

The configuration is a two-dimensional planar hypersonic flow over a circular
cylinder of radius \(R\).
The incoming stream is aligned with the symmetry plane, and the upper half-plane
is used for the analysis and visualization. The cylinder geometry is specified
so that the leftmost point of the body is located at \(x=0\), giving \(x_c=R\)
and \(y_c=0\) for the cylinder centre. The freestream Knudsen number is defined
using the cylinder diameter \(D=2R\),
\begin{equation}
    Kn_\infty=\frac{\lambda_\infty}{D},
    \label{eq:kn_definition}
\end{equation}
where \(\lambda_\infty\) is the freestream mean free path associated with the
corresponding DSMC inflow state. The primary Knudsen-number sweep is performed
at \(M_\infty=10\), while an additional Mach-number sweep at
\(Kn_\infty=0.01\) is used to separate rarefaction effects from changes in
shock strength. Both argon and nitrogen are considered, so that monatomic and
rotationally relaxing gas responses can be compared. The geometry
is a two-dimensional planar cylinder problem. All standoff, thickness and modal quantities therefore
refer to this two-dimensional cylinder configuration.

The DSMC simulations were performed using Bird's DS2V direct simulation Monte
Carlo solver. The hypersonic cylinder case corresponds to a two-dimensional
circular cylinder of diameter \(D=0.3048~\mathrm{m}\), freestream temperature
\(T_\infty=200~\mathrm{K}\), and a fully diffuse cylinder wall held at
\(T_w=500~\mathrm{K}\). Gas--gas collisions were modelled using the
variable-hard-sphere (VHS) model. For argon, the molecular mass is
\(m=6.63\times10^{-26}~\mathrm{kg}\), and the VHS reference parameters are
\(T_{\mathrm{ref}}=273~\mathrm{K}\),
\(d_{\mathrm{ref}}=4.17\times10^{-10}~\mathrm{m}\), and \(\omega=0.81\). For
nitrogen, the molecular mass is \(m=4.65\times10^{-26}~\mathrm{kg}\), with the
same reference temperature and reference diameter,
\(T_{\mathrm{ref}}=273~\mathrm{K}\) and
\(d_{\mathrm{ref}}=4.17\times10^{-10}~\mathrm{m}\), and \(\omega=0.74\). The
nitrogen calculations use the standard DS2V rotational nonequilibrium treatment
for a diatomic gas; vibrational excitation and chemistry are not included in
the benchmark fields used here. The computational domain was discretized using
\(194\times100\) divisions, and approximately \(1.5\times10^{6}\) simulator
particles were used. During the simulations, the collision cells were
continuously adapted to maintain an average of about 20 particles per cell,
providing adequate resolution of the collision statistics throughout the
domain. The DSMC data and the corresponding cylinder benchmark were previously
validated against Bird's reference solution and related DSMC calculations
\citep{Goshayeshi2015SBTTAS,Goshayeshi2015SBT}. The present post-processing uses
the time-averaged DS2V macroscopic fields. The solver and the underlying
cylinder benchmark fields have already been validated in the cited DSMC studies;
the additional checks in the present paper therefore target the extracted
shock-layer metrics, registration procedure and modal conclusions rather than
constituting a new solver-verification exercise. Independent block-averaged
confidence intervals were not available for all cases, so the shock-layer
extraction uses median upstream/downstream estimates, a fixed ray sector and
comparative POD metrics to reduce sensitivity to local DSMC sampling scatter.
The robustness analysis in \appref{app:robustness} further varies the
extraction settings to test whether the modal conclusions depend on reasonable
post-processing choices.

\begin{table}
\centering
\caption{Summary of the simulation and post-processing parameters used in the
present analysis. The freestream density is varied consistently with the target
\(Kn_\infty=\lambda_\infty/D\); all cases use the same cylinder diameter,
wall condition, VHS gas model and ray-processing workflow.}
\label{tab:case_summary}
\begin{tabular}{ll}
\toprule
Quantity & Value or definition \\
\midrule
Geometry & Two-dimensional planar circular cylinder, \(D=0.3048~\mathrm{m}\) \\
Wall condition & Fully diffuse wall, \(T_w=500~\mathrm{K}\) \\
Freestream temperature & \(T_\infty=200~\mathrm{K}\) \\
Knudsen sweep & \(M_\infty=10\), \(Kn_\infty\approx0.01\)--1 \\
Mach sweep & \(Kn_\infty=0.01\), \(M_\infty=5\)--15 \\
Gases & Ar and \(N_2\), VHS collisions \\
Nitrogen internal energy & Rotational nonequilibrium in DS2V; no vibration/chemistry \\
Grid and particles & \(194\times100\) divisions, about \(1.5\times10^6\) simulator particles \\
Adaptive cells & Approximately 20 simulator particles per collision cell \\
Ray sector & \(100^\circ\leq\theta\leq180^\circ\), 65 body-normal rays \\
Ray and attached grids & 900 ray points; 260 points in \(-1\leq\xi_q\leq4\) \\
Main marker & Maximum radial density-gradient station for \(\rho\) registration \\
\bottomrule
\end{tabular}
\end{table}

The wall temperature is therefore higher than the freestream temperature
(\(T_w/T_\infty=2.5\)), but it is still below the translational temperature
generated inside the hypersonic post-shock layer. The wall boundary condition
should consequently be interpreted as a relatively warm diffuse wall compared
with the incoming gas, but as a weaker thermal sink for the shock-heated gas
than an isothermal wall held at \(T_\infty\). This distinction is important for
the interpretation of the thermal fields: the density-compression layer is
mainly controlled by the incoming momentum flux and rarefaction, whereas the
near-wall translational and rotational temperature fields are also influenced
by the finite wall temperature and by gas--surface energy accommodation.

For clarity, the shock-attached notation used below is defined here. The radial
distance from the cylinder centre is \(r\), and \(s=r-R\) is the distance
measured from the wall along a body-normal ray. The polar angle \(\theta\) is
measured from the positive \(x\)-axis, \(q\) denotes a generic macroscopic
field, \(q_{\mathrm{up}}\) and \(q_{\mathrm{down}}\) are robust upstream and
downstream ray averages, \(r_q\) or \(s_q\) is the location of the maximum
radial gradient of \(q\), \(\delta_q\) is the corresponding gradient-based
thickness, and \(\xi_q\) is the variable-attached coordinate. The symbols
\(E_i\) and \(C_i\) denote, respectively, the fractional POD energy of mode
\(i\) and the cumulative energy through mode \(i\). These definitions are
repeated in the equations below when they enter the analysis.

\subsection{Shock detection and validation}
\label{sec:shock_detection_validation}

The present analysis requires a reproducible definition of the bow-shock
location in DSMC fields where the compression layer has a finite kinetic
thickness rather than a discontinuous jump. For cases in which a distinct edge
exists, we define the shock centre as a density-gradient ridge and compare it
with the schlieren-based shock-wave detection (SWD) method of
\citet{Akhlaghi2017SWD}, later applied to rarefied shock polars over cylinders
by \citet{Akhlaghi2021ShockPolar}.

Let \((x_c,y_c)\) and \(R\) denote the centre and radius of the circular
cylinder. For each polar angle \(\theta\), the DSMC field is sampled along the
ray
\begin{equation}
    \boldsymbol{x}(r;\theta)
    =
    (x_c,y_c)
    +
    r(\cos\theta,\sin\theta),
    \qquad r>R .
    \label{eq:ray_coordinate}
\end{equation}
The low-\(Kn_\infty\) shock-centre radius is defined as the first significant
upstream density-gradient peak outside a near-wall exclusion region,
\begin{equation}
    r_s(\theta)
    =
    \arg\max_{r\in\mathcal{I}_s(\theta)}
    \left|
    \frac{\partial \rho}{\partial r}
    \left(\boldsymbol{x}(r;\theta)\right)
    \right|,
    \label{eq:shock_ray_definition}
\end{equation}
where \(\mathcal{I}_s\) is restricted to the upstream compression layer and
excludes points too close to the cylinder surface. The corresponding local
standoff distance is
\begin{equation}
    \Delta_s(\theta)=r_s(\theta)-R .
    \label{eq:standoff_definition}
\end{equation}

For each macroscopic variable \(q\), we define a variable-specific gradient
thickness,
\begin{equation}
    \delta_q(\theta)
    =
    \frac{
    \left|q_{\mathrm{down}}(\theta)-q_{\mathrm{up}}(\theta)\right|
    }{
    \max_{r\in\mathcal{I}_s(\theta)}
    \left|\partial q/\partial r\right|
    },
    \label{eq:variable_thickness}
\end{equation}
where \(q_{\mathrm{up}}\) and \(q_{\mathrm{down}}\) are robust upstream and
downstream averages along the same ray. The associated attached coordinate is
\begin{equation}
    \xi_q=\frac{r-r_q}{\delta_q},
    \label{eq:variable_attached_coordinate}
\end{equation}
where \(r_q\) is the location of the maximum radial gradient of \(q\). This
definition is used for density, Mach number, pressure, translational
temperature and, for nitrogen, rotational temperature.

The ray extraction and registration were implemented with fixed numerical
settings for all cases. The upstream shock sector was sampled with 65 uniformly
spaced body-normal rays over \(100^\circ\leq \theta \leq 180^\circ\), where
\(180^\circ\) corresponds to the stagnation line in front of the cylinder. Each
ray was sampled using 900 points from the wall to either the computational-domain
boundary or \(8R\), whichever was reached first. The scattered DSMC cell-centred
data were evaluated on the ray by piecewise-linear interpolation over a Delaunay
triangulation of the original \((x,y)\) points. The profile was mildly smoothed
only for locating the maximum-gradient station and the gradient thickness; the
registered field values themselves were obtained from the normalized ray profile
by one-dimensional linear interpolation onto the common grid
\(-1\leq \xi_q\leq 4\) with 260 points. Thus the POD is performed on the same
attached grid for every snapshot, while the original invalid or unsupported
regions remain masked.

The search interval \(\mathcal{I}_s(\theta)\) excludes the near-wall region
\(s<0.015R\) and the far end of the sampled ray \(s>0.94s_{\max}\), so that wall
jumps and domain-boundary plateaux are not selected as shock-layer gradients.
The downstream value \(q_{\mathrm{down}}\) is the median of the smoothed profile
between 3\% and 23\% of the valid ray length outside the wall, while
\(q_{\mathrm{up}}\) is the median over the farthest 15\% of the valid ray. These
median averages make the jump estimate insensitive to individual noisy DSMC
cells. The precise numerical modal energies can depend weakly on the selected angular
sector, attached-grid resolution and interpolation procedure.
For this reason, the quoted energies are interpreted as diagnostics for the
stated shock-attached representation rather than as universal constants. The
central comparison is more robust than the individual numbers because all
variables are processed with the same ray sector, interpolation, masking and
POD procedure. \appref{app:robustness} repeats the extraction with
alternative angular sectors, ray numbers, attached-grid resolutions, smoothing
widths and common-support thresholds.

The SWD comparison is used only as a validation in cases where a narrow
schlieren-like edge is present. In the SWD method, the density field is first
converted to a numerical schlieren image, and a Roberts-cross edge detector is
then applied to obtain a local shock-response function. Along each local
row or sampling direction, the response in the shock neighbourhood is
approximated by a Gaussian profile,
\begin{equation}
    G_i(x)
    =
    B_{1,i}
    \exp\left[
    -\frac{(x-B_{2,i})^2}{2B_{3,i}^{\,2}}
    \right],
    \label{eq:akhlaghi_gaussian}
\end{equation}
where \(i\) denotes the local sampling row, \(G_i(x)\) is the fitted
schlieren-response profile, \(B_{1,i}\) is the peak response amplitude,
\(B_{2,i}\) is the shock-centre location, and \(B_{3,i}\) is the Gaussian
width parameter associated with the finite shock thickness. The normalized
within-shock coordinate is then defined as
\begin{equation}
    \sigma_i(x)=\frac{G_i(x)}{B_{1,i}} .
    \label{eq:akhlaghi_sigma}
\end{equation}
With this definition, \(\sigma_i=1\) occurs at \(x=B_{2,i}\) and denotes
the SWD shock centre, while lower values of \(\sigma_i\) identify locations
away from the centre of the finite-thickness shock-response profile. In the
present comparison, \(\sigma_i=0.1\) is used to mark the aft or post-shock
side of the fitted shock layer.

\begin{figure}
    \centering
    \includegraphics[
        width=0.75\textwidth,
        trim={0pt 0pt 0pt 0pt},
        clip
    ]{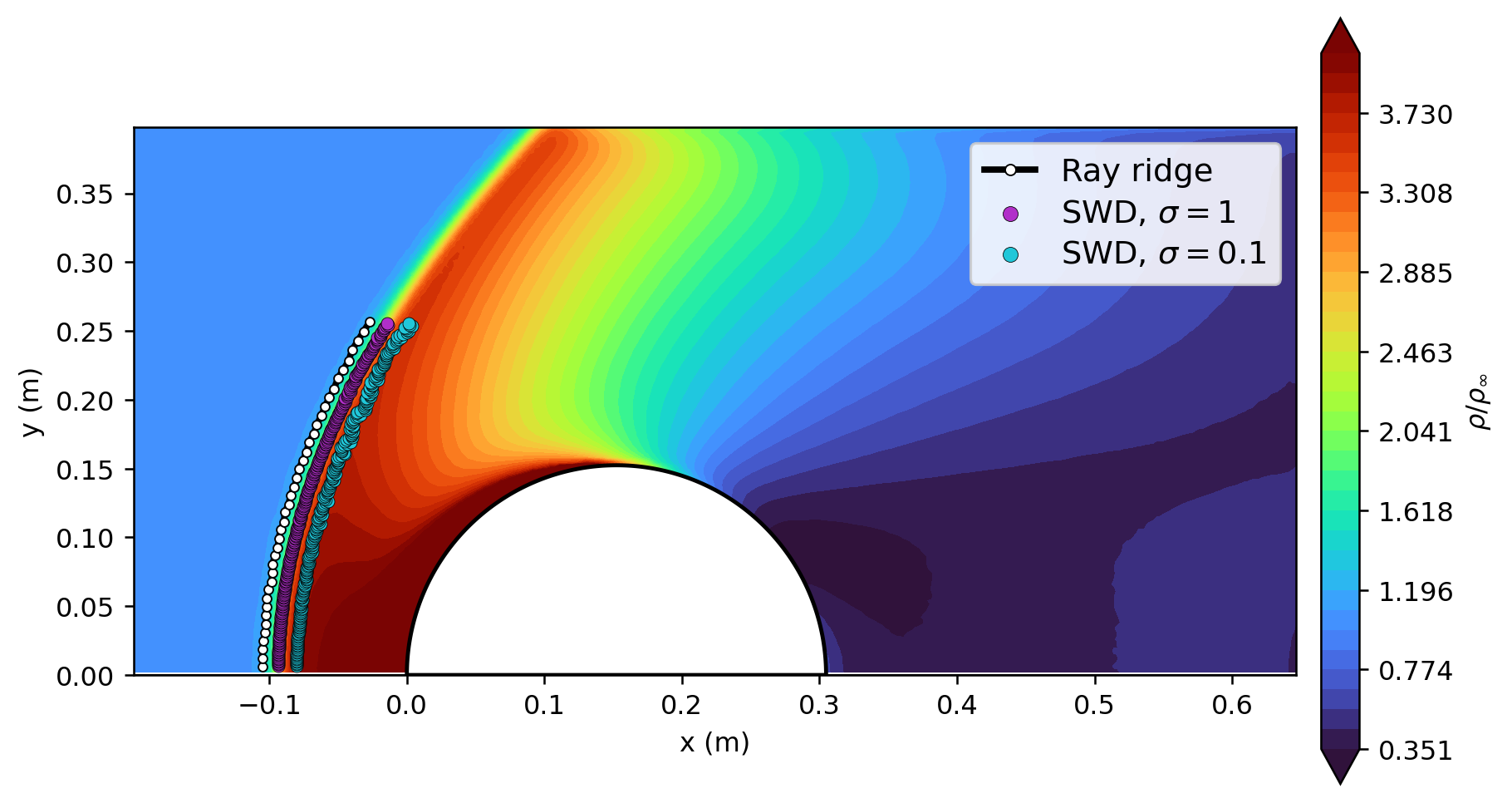}
    \caption{
    Validation of the low-\(Kn_\infty\) shock-front extraction for the
    representative argon case at \(Kn_\infty=0.01\) and \(M_\infty=10\).
    The black curve with white markers denotes the ray-based
    density-gradient ridge used in the present analysis. The purple points
    show the Akhlaghi et al. SWD shock-centre locations \((\sigma=1)\), while
    the cyan points show the corresponding aft-shock locations
    \((\sigma=0.1)\). The close agreement between the ray ridge and the SWD
    shock-centre points supports the use of the ray-based extraction in the
    low-rarefaction regime where a distinct shock edge exists.
    }
    \label{fig:shock_detection_validation}
\end{figure}

Figure~\ref{fig:shock_detection_validation} shows a representative validation
case at \(Kn_\infty=0.01\) and \(M_\infty=10\). The Akhlaghi et al. SWD
shock-centre points agree closely with the ray-based density-gradient ridge,
while the \(\sigma\simeq0.1\) points lie on the downstream side of the
finite-thickness shock layer. At larger Knudsen numbers, the bow shock becomes
a broad kinetic compression layer rather than a narrow image edge; in that
regime, the present work uses layer-based measures and does not impose a unique
visual shock line.

\section{Results and discussion}
\label{sec:results_discussion}

\subsection{Rarefaction-induced bow-shock inflation}
\label{sec:inflation}

\begin{figure}
    \centering
    \begin{subfigure}{0.98\textwidth}
        \centering
        \includegraphics[
            width=\textwidth,
            trim={0pt 0pt 0pt 55pt},
            clip
        ]{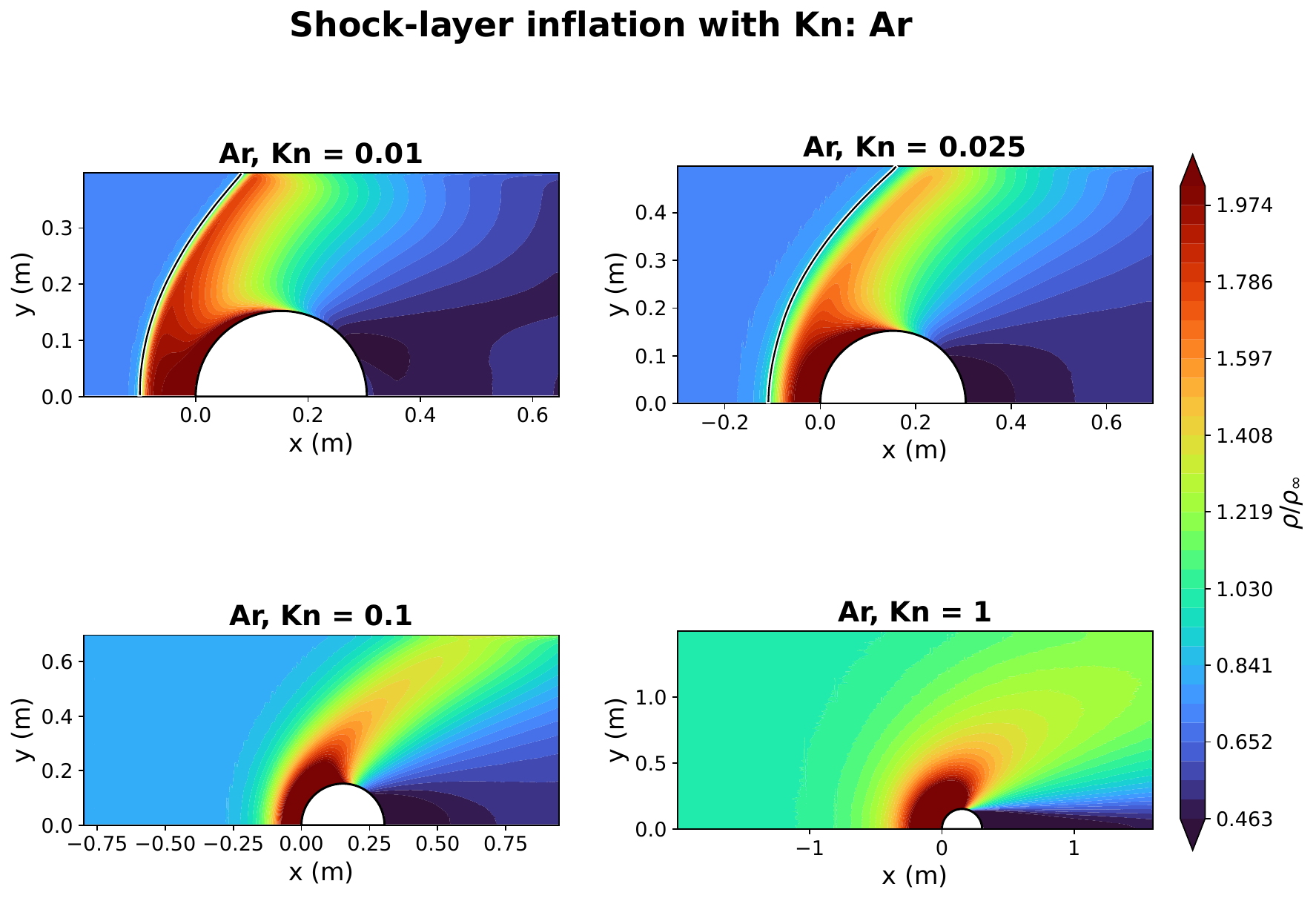}
        \caption{Argon.}
        \label{fig:kn_rho_ar}
    \end{subfigure}

    \vspace{0.35em}

    \begin{subfigure}{0.98\textwidth}
        \centering
        \includegraphics[
            width=\textwidth,
            trim={0pt 0pt 0pt 55pt},
            clip
        ]{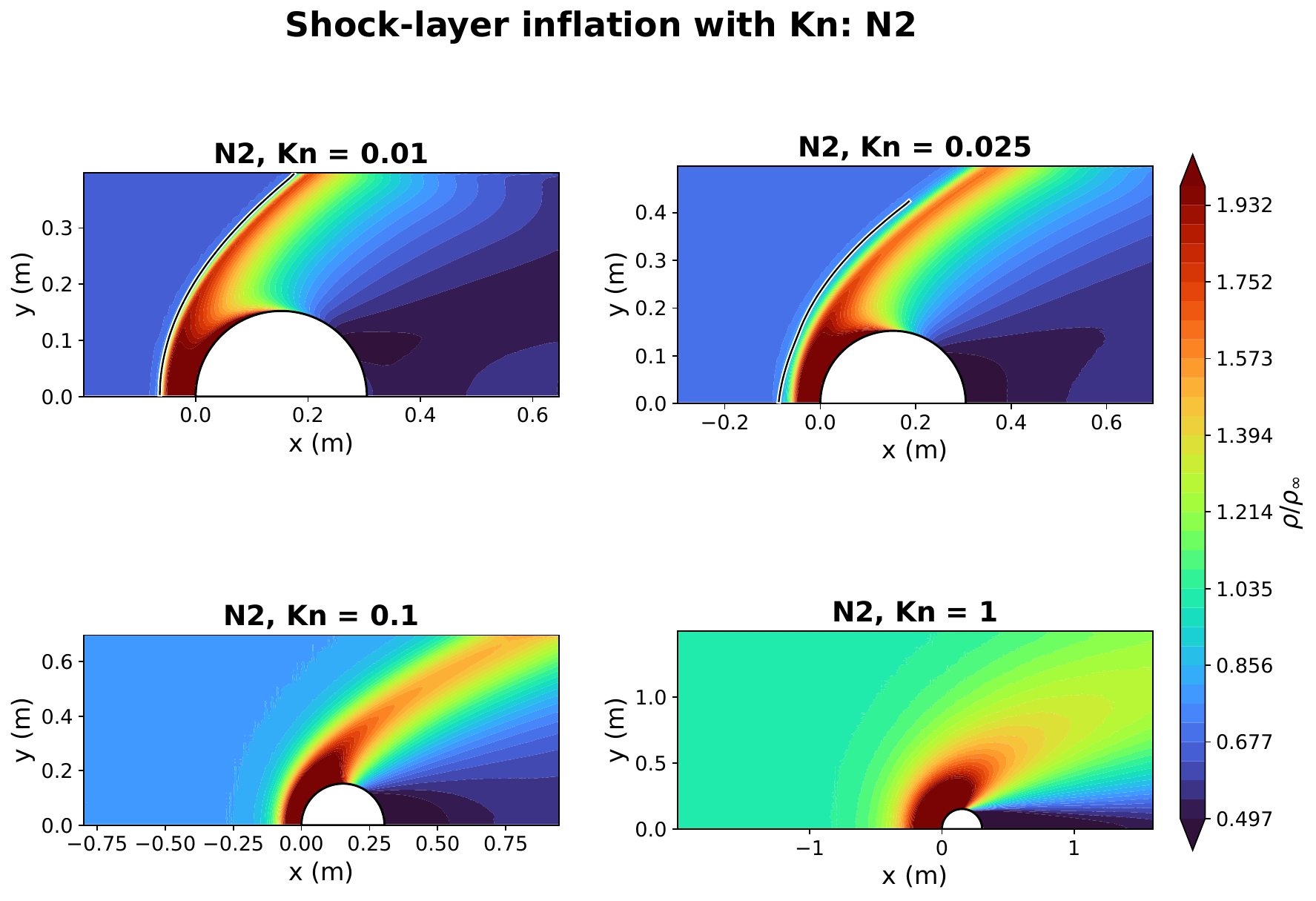}
        \caption{Nitrogen.}
        \label{fig:kn_rho_n2}
    \end{subfigure}

    \caption{
    Rarefaction-induced inflation of the hypersonic bow-shock layer over a
    circular cylinder at \(M_\infty=10\). The panels show the normalized density
    field \(\rho/\rho_\infty\) for increasing \(Kn_\infty\). For the low-Kn
    cases, where a distinct compression-front ridge exists, the detected front
    is overlaid. At larger \(Kn_\infty\), the bow shock becomes a broad kinetic
    compression layer and no unique shock-line overlay is imposed. The colour scale is kept common within each gas to emphasize the growth of the
compressed region. Since \(\rho/\rho_\infty=1\) lies inside the colour range
rather than at its lower bound, the undisturbed upstream state is not
necessarily represented by the darkest blue colour. The upstream normalization
was checked directly from the far-field DSMC samples for each case.
    }
    \label{fig:kn_density_inflation}
\end{figure}

Figure~\ref{fig:kn_density_inflation} shows the density field for the
Knudsen-number sweep at fixed \(M_\infty=10\). At \(Kn_\infty=0.01\) and
\(0.025\), both gases exhibit a distinct bow-shock front, and the detected
front is overlaid for reference. As \(Kn_\infty\) increases to \(0.1\) and
\(1\), the sharp front is replaced by a spatially extended compression layer.
The high-\(Kn_\infty\) panels are therefore interpreted as shock-layer
inflation rather than displacement of a single discontinuity.

The argon and nitrogen fields show the same qualitative progression: upstream
broadening of the density-compression region, weakening of the sharp density
ridge, and increased interaction between the compression layer and the near-body
region. The similarity of the density fields indicates that rarefaction is the
dominant control parameter for the gross inflation of the bow-shock layer,
whereas differences between argon and nitrogen become more important in the
thermal and internal-energy relaxation fields analysed below.

\subsection{Mach-number variation at fixed rarefaction}
\label{sec:mach_sweep}

\begin{figure}
    \centering
    \begin{subfigure}{0.98\textwidth}
        \centering
        \includegraphics[
            width=\textwidth,
            trim={0pt 0pt 0pt 55pt},
            clip
        ]{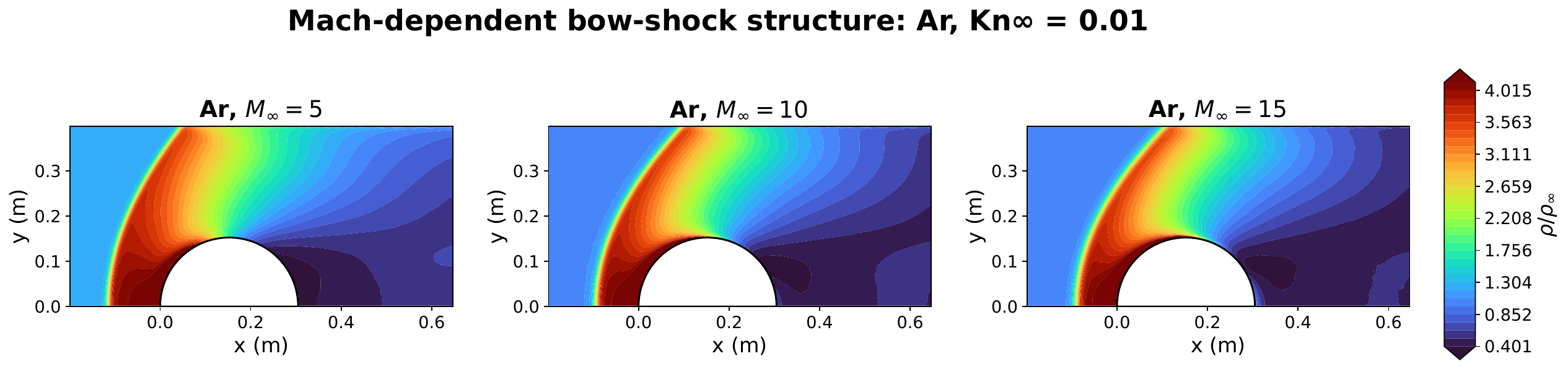}
        \caption{Argon.}
        \label{fig:mach_rho_ar}
    \end{subfigure}

    \vspace{0.35em}

    \begin{subfigure}{0.98\textwidth}
        \centering
        \includegraphics[
            width=\textwidth,
            trim={0pt 0pt 0pt 55pt},
            clip
        ]{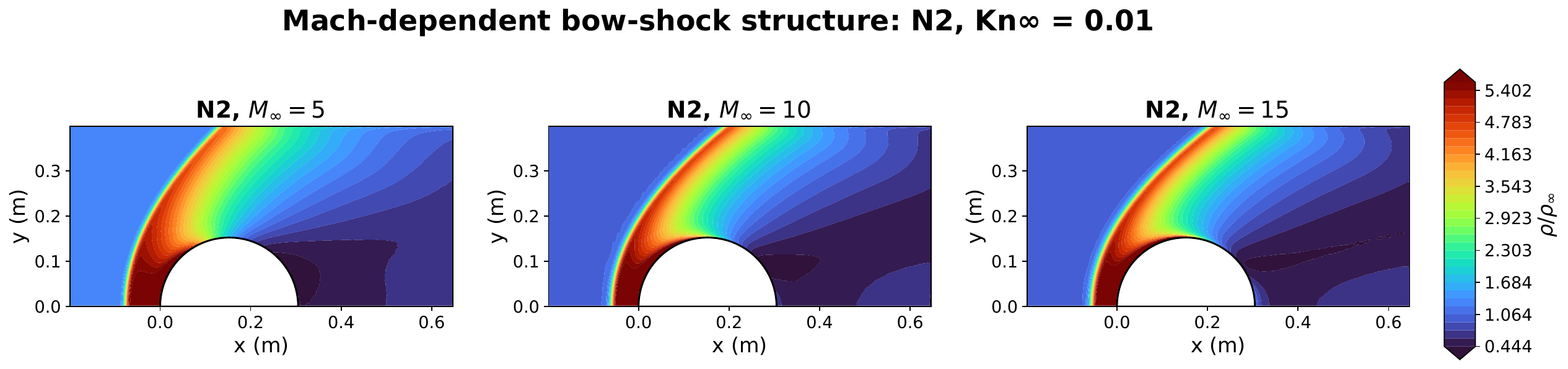}
        \caption{Nitrogen.}
        \label{fig:mach_rho_n2}
    \end{subfigure}

    \caption{
Mach-number dependence of the bow-shock structure at fixed
\(Kn_\infty=0.01\). Increasing \(M_\infty\) strengthens the compression and
moves the upstream bow-shock branch closer to the cylinder, consistent with the
classical decrease of blunt-body shock standoff with Mach number. Unlike the
Knudsen-number sweep, the shock layer remains comparatively sharp and coherent,
showing that changing shock strength at low rarefaction does not produce the
same kinetic delocalization as increasing \(Kn_\infty\). Nitrogen exhibits a
somewhat broader post-shock relaxation region because part of the
shock-generated translational energy is redistributed into rotational modes.
}
    \label{fig:mach_density_sweep}
\end{figure}

Figure~\ref{fig:mach_density_sweep} shows the Mach-number dependence of the
density field at fixed \(Kn_\infty=0.01\). At this low rarefaction level, the
flow still retains a relatively continuum-like detached bow-shock structure.
Increasing \(M_\infty\) strengthens the upstream compression and increases the
normal-shock density jump. As a result, the gas can be decelerated and
compressed over a shorter distance ahead of the cylinder, and the upstream
branch of the bow shock moves closer to the body. Equivalently, the shock
standoff distance decreases as the post-shock density and pressure rise become
larger: a thinner compressed layer is sufficient to turn the flow around the
blunt body. This trend is consistent with the classical continuum behaviour of
detached shocks over blunt bodies, where the shock standoff distance decreases
with increasing Mach number and approaches a high-Mach-number asymptote once
the normal-shock compression ratio becomes nearly saturated.

The Mach-number sweep should therefore be interpreted differently from the
Knudsen-number sweep. Varying \(M_\infty\) at fixed low \(Kn_\infty\) mainly
changes the strength, curvature and standoff distance of an otherwise coherent
bow shock. In contrast, increasing \(Kn_\infty\) changes the kinetic character
of the layer itself: the mean free path becomes comparable to the shock-layer
thickness, collisions are no longer frequent enough to maintain a narrow
compression front, and the shock is replaced by a broad non-equilibrium
compression layer. Thus, the Mach sweep produces a mostly geometric
displacement and strengthening of the shock, whereas the Knudsen-number sweep
produces a qualitative loss of a unique shock-front representation.

The argon and nitrogen fields follow the same leading-order Mach trend, but
their post-shock relaxation structures differ. In monatomic argon, the shock
energy is accommodated only through translational degrees of freedom, so the
density and translational-temperature adjustment remain more tightly coupled.
In nitrogen, part of the shock-generated translational energy is transferred
into rotational modes over a finite relaxation distance. This internal-energy
exchange broadens the thermal relaxation region behind the density-compression
front and makes the nitrogen shock layer less single-scale than the argon
layer. Consequently, the gas dependence is not primarily a change in the
existence of the bow shock at \(Kn_\infty=0.01\), but a change in how the
post-shock non-equilibrium structure relaxes downstream of the compression
front.

\subsection{Quantitative shock-layer metrics}
\label{sec:metrics}

The density-contour visualizations establish the qualitative inflation of the
bow-shock layer, but a quantitative comparison requires metrics that do not rely
on manually drawing a single shock line in the high-Knudsen-number cases. We
therefore extract the metrics directly from one-dimensional profiles sampled
along body-normal rays. For each ray, the density-front location is defined as
the position of the maximum radial density gradient. At low \(Kn_\infty\), this
location is interpreted as the shock-front position; at higher \(Kn_\infty\),
where no unique discontinuity exists, it is interpreted as the characteristic
location of the strongest compression within the broad kinetic layer. The
effective density thickness is computed as
\begin{equation}
    \delta_\rho
    =
    \frac{|\rho_{\mathrm{down}}-\rho_{\mathrm{up}}|}
    {\max |\partial \rho/\partial r|},
    \label{eq:delta_rho_metric}
\end{equation}
and is supplemented by a robust 10--90 thickness extracted from the normalized
profile. These quantities are obtained from the local ray profiles and then
summarized by the median over the upstream angular sector. Thus the metrics used
below do not require imposing a visible shock curve on the diffuse
\(Kn_\infty=0.1\) and \(1\) fields.

This distinction is essential for the high-\(Kn_\infty\) cases. At
\(Kn_\infty=0.1\) and especially at \(Kn_\infty=1\), the maximum-gradient point
is not treated as a discontinuity or as the unique physical shock location; it
is only a reproducible marker of the strongest compression within a broad
kinetic layer. The standoff and thickness metrics therefore describe the
inflated compression layer, not a continuum shock surface.

\begin{figure}
    \centering
    \includegraphics[
        width=0.90\textwidth,
        trim={0pt 0pt 0pt 30pt},
        clip
    ]{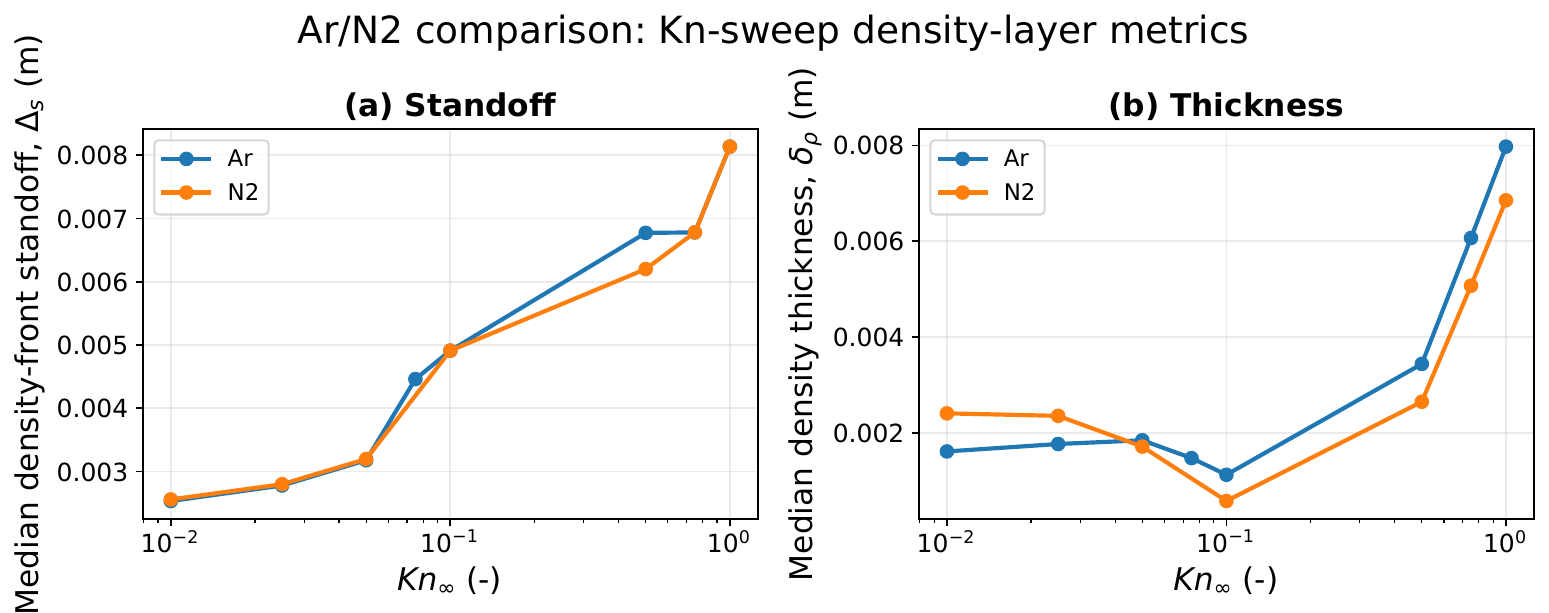}
    \caption{
Density-based shock-layer metrics for the Knudsen-number sweep at
\(M_\infty=10\). Panel (a) shows the median density-front standoff and panel
(b) shows the effective density thickness
\(\delta_\rho=|\rho_{\mathrm{down}}-\rho_{\mathrm{up}}|/
\max|\partial\rho/\partial r|\). Because \(M_\infty\) is fixed, the continuum normal-shock density ratio
provides an approximately fixed low-rarefaction reference compression scale for
each gas; the increase in standoff therefore reflects kinetic rarefaction rather
than a change in inviscid shock-strength reference. The thickness has
a shallow minimum near \(Kn_\infty\simeq0.1\) because the sampled density jump
and the peak density gradient vary in opposite ways. At higher \(Kn_\infty\),
collisional localization breaks down and the density change is distributed over
a mean-free-path-controlled compression layer.
}
    \label{fig:kn_density_metrics}
\end{figure}

Figure~\ref{fig:kn_density_metrics} quantifies the rarefaction-induced
evolution of the density-compression layer. The key point is that this sweep
is performed at fixed \(M_\infty=10\). In an inviscid hypersonic blunt-body
picture, the leading-order standoff is organized primarily by the inverse
normal-shock density ratio
\begin{equation}
    \epsilon
    =
    \frac{\rho_\infty}{\rho_s}
    =
    \frac{\gamma-1+2/M_\infty^2}{\gamma+1},
    \label{eq:epsilon_kn_discussion}
\end{equation}
where \(\rho_s\) is the density immediately behind a normal shock. Hornung and
co-workers showed that detached-shock standoff distances for spheres, cones,
wedges and circular cylinders are strongly organized by this inverse post-shock
density ratio, with extensions to non-equilibrium blunt-body flows through an
effective density-ratio interpretation
\citep{Hornung1972Nonequilibrium,Hornung2019SphereCone,Hornung2021WedgeCylinder}.
In the present Knudsen-number sweep, however, \(\epsilon\) is approximately
fixed for each gas. Therefore the systematic increase of the measured
density-front standoff in figure~\ref{fig:kn_density_metrics}(a) cannot be a
normal-shock compression-ratio effect. It is a kinetic displacement: as the mean
free path increases, the incoming directed kinetic energy is not converted into
a localized post-shock compression over the same short collisional distance, and
the strongest density-gradient response moves upstream.

A schematic way to express this distinction, used here as an interpretive decomposition rather than a fitted law, is
\begin{equation}
    \frac{\Delta_s}{R}
    \simeq
    \Phi_{\mathrm{inv}}(\epsilon)
    +
    \Phi_{\mathrm{kin}}(Kn_s),
    \label{eq:standoff_decomposition}
\end{equation}
where \(R\) is the cylinder radius, \(\Phi_{\mathrm{inv}}\) denotes the
compression-ratio contribution, and \(Kn_s=\lambda_s/R\) is a representative
post-shock or shock-layer Knudsen number. For the Mach sweep,
\(\Phi_{\mathrm{inv}}\) changes because \(\epsilon\) changes. For the
Knudsen-number sweep, \(\Phi_{\mathrm{inv}}\) is nearly fixed, while
\(\Phi_{\mathrm{kin}}\) grows with the mean free path. This explains why the
Knudsen-number sweep inflates the compression layer even though the nominal
shock strength is unchanged.

This interpretation is consistent with Hornung's analysis of non-equilibrium
nitrogen flow over spheres and circular cylinders. There, finite-rate
relaxation introduces a length scale behind the shock; as the relaxation rate
decreases, the relaxation zone grows from a thin region near the shock into a
layer occupying a substantial part of the shock layer. The present flow is
non-dissociating, but the analogy is direct: increasing \(Kn_\infty\) reduces
the collision rate responsible for local thermalization, so density and
temperature adjust over a larger kinetic region. The high-\(Kn_\infty\) panels
should therefore be interpreted as finite-thickness kinetic compression layers,
not as ordinary continuum shocks merely shifted away from the wall.

The thickness trend in figure~\ref{fig:kn_density_metrics}(b) is governed by
the same competition. Because \(\delta_\rho\) is defined by
\eqref{eq:delta_rho_metric}, it depends on both the density jump sampled across
the compression layer and the peak density-gradient strength. Between
\(Kn_\infty=0.01\) and \(Kn_\infty\simeq0.1\), the compression front remains
organized enough to maintain a strong gradient, while the effective density
jump decreases as the post-shock state departs from the continuum-like
normal-shock value. The ratio can therefore decrease, producing the observed
shallow minimum in \(\delta_\rho\). At larger \(Kn_\infty\), the balance
reverses: the peak gradient weakens, the compression is distributed over a
larger kinetic distance, and the effective thickness becomes controlled by the
local mean free path,
\begin{equation}
    \delta_\rho/R = O(Kn_s),
    \qquad Kn_s=\lambda_s/R .
    \label{eq:delta_rho_high_kn_scaling}
\end{equation}
Thus the increase of \(\delta_\rho\) at high \(Kn_\infty\) reflects breakdown of
collisional localization rather than a change in inviscid shock compression.

The argon and nitrogen curves follow the same leading-order trend, confirming
that the density-layer inflation is primarily controlled by rarefaction. The
remaining gas dependence is physically meaningful but secondary. In argon, all
post-shock energy remains translational, so the density and translational
temperature adjustments are tightly coupled. In nitrogen, rotational relaxation
introduces an additional thermodynamic length scale behind the density front,
modifying the detailed post-shock pressure and density adjustment without
changing the dominant kinetic mechanism. Therefore
figure~\ref{fig:kn_density_metrics} supports a Hornung-like interpretation in
the low-\(Kn_\infty\) limit, but shows that increasing \(Kn_\infty\) adds a
mean-free-path-controlled inflation term absent from inviscid perfect-gas
standoff correlations.

\begin{figure}
    \centering
    \includegraphics[
        width=0.90\textwidth,
        trim={0pt 0pt 0pt 30pt},
        clip
    ]{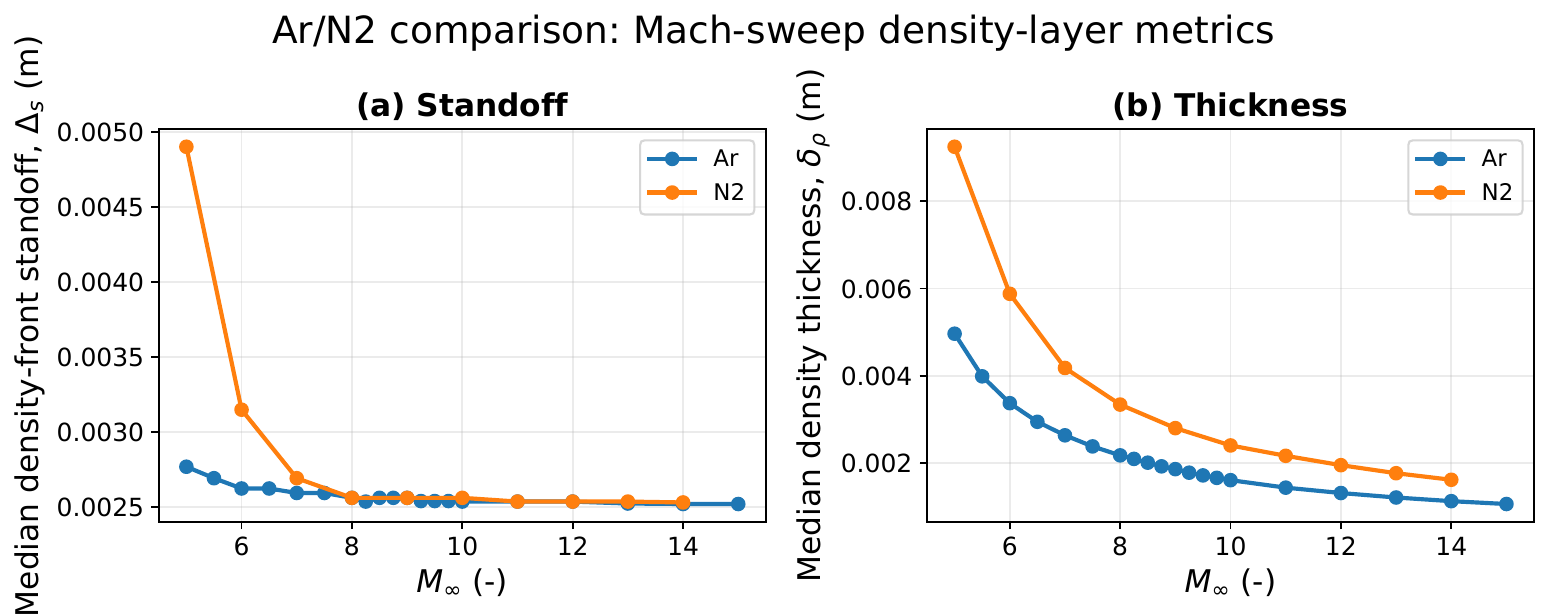}
    \caption{
    Density-based shock-layer metrics for the Mach-number sweep at fixed
    \(Kn_\infty=0.01\). Increasing \(M_\infty\) reduces the inverse
    normal-shock density ratio and moves the density-compression layer closer
    to the cylinder. The argon standoff approaches a high-Mach plateau because
    the normal-shock density ratio is already close to its strong-shock limit.
    The effective density thickness decreases because the peak density
    gradient strengthens faster than the sampled density jump grows. Nitrogen
    follows the same leading compression trend, with secondary differences
    associated with rotational relaxation behind the density front.
    }
    \label{fig:mach_density_metrics}
\end{figure}

Figure~\ref{fig:mach_density_metrics} gives the corresponding density-based
metrics for the Mach-number sweep at fixed \(Kn_\infty=0.01\). This sweep is
the complementary limit to figure~\ref{fig:kn_density_metrics}: the flow
remains at low rarefaction, so the kinetic contribution to the standoff is
comparatively small, but the inverse normal-shock density ratio
\(\epsilon\), defined in \eqref{eq:epsilon_kn_discussion}, changes with
\(M_\infty\). For a circular cylinder, the inviscid Hornung correlation can be
written approximately as
\begin{equation}
    \frac{\Delta}{R}
    \simeq
    2.14\,\epsilon\left(1+\frac{\epsilon}{2}\right),
    \label{eq:hornung_cylinder_standoff}
\end{equation}
with the older near-linear form \(\Delta/R\simeq2.32\epsilon\) giving a similar
strong-shock trend
\citep{Hornung1972Nonequilibrium,Hornung2021WedgeCylinder}. Although the
present DSMC metric \(\Delta_s\) is the maximum-density-gradient location
rather than the inviscid sonic-line standoff, the same compression-ratio
scaling explains why increasing \(M_\infty\) moves the density-compression
layer closer to the cylinder.

This scaling also explains the argon plateau in panel (a). For monatomic argon,
\(\gamma=5/3\), and the normal-shock density ratio tends to the finite
strong-shock limit \((\gamma+1)/(\gamma-1)=4\), or
\(\epsilon\rightarrow0.25\). The corresponding compression ratios at
\(M_\infty=5,10\) and 15 are approximately 3.57, 3.88 and 3.95. Most of the
available density compression has therefore already been achieved by
\(M_\infty=10\). Increasing the Mach number from 10 to 15 still raises the
kinetic energy and post-shock pressure, but it only weakly changes the density
ratio controlling the leading geometric standoff; the argon curve therefore
approaches a high-Mach asymptote.

The nitrogen standoff follows the same leading-order compression argument but
need not coincide with the argon curve. A calorically perfect diatomic estimate
with \(\gamma\simeq1.4\) gives a strong-shock density-ratio limit of 6, which
would favour a smaller inviscid standoff for the same body scale. In the DSMC
nitrogen cases, however, finite rotational relaxation modifies the downstream
pressure and temperature adjustment behind the density front. This is analogous
to Hornung's non-equilibrium nitrogen analysis, where density patterns over
blunt bodies are sensitive to finite-rate relaxation; here the additional
relaxation mechanism is rotational rather than chemical.

Panel (b) shows the corresponding effective density thickness, defined in
\eqref{eq:delta_rho_metric}. Its decrease with \(M_\infty\) is not simply a
consequence of the density jump increasing. At fixed \(Kn_\infty=0.01\), the
flow remains sufficiently collisional that stronger Mach-number compression is
localized into a sharper density-gradient ridge. The density jump increases
toward its strong-shock limit, but the peak gradient increases more rapidly
than the jump itself, so the effective thickness decreases. This is opposite to
the high-\(Kn_\infty\) behaviour in figure~\ref{fig:kn_density_metrics}, where
rarefaction weakens the peak gradient and spreads the compression over a larger
kinetic length.

Thus the four curves in figure~\ref{fig:mach_density_metrics} support a
consistent interpretation: at fixed low \(Kn_\infty\), increasing
\(M_\infty\) primarily decreases \(\epsilon\), moves the density-compression
layer closer to the body and sharpens the density-gradient thickness. The
behaviour is Hornung-like in the sense that the gross standoff is governed by
post-shock compression, while the gas-dependent deviations reveal relaxation
physics absent from a purely inviscid perfect-gas correlation. This is
fundamentally different from the Knudsen-number sweep, where the loss of
collisional localization, rather than the normal-shock compression ratio,
controls the growth of the shock-layer thickness.

The same trends are summarized in \appref{app:empirical_scaling} using
empirical diagnostic fits of the extracted standoff and thickness metrics. Those
fits are not introduced as universal shock correlations; they are used only as
compact summaries of the DSMC-derived trends. In particular, the Knudsen-sweep
standoff fit highlights the monotonic kinetic displacement of the strongest
compression marker, whereas the density-thickness fit emphasizes the shallow
intermediate-
\(Kn_\infty\) minimum followed by high-
\(Kn_\infty\) broadening.

\begin{figure}
    \centering
    \includegraphics[
        width=0.98\textwidth,
        trim={0pt 0pt 0pt 30pt},
        clip
    ]{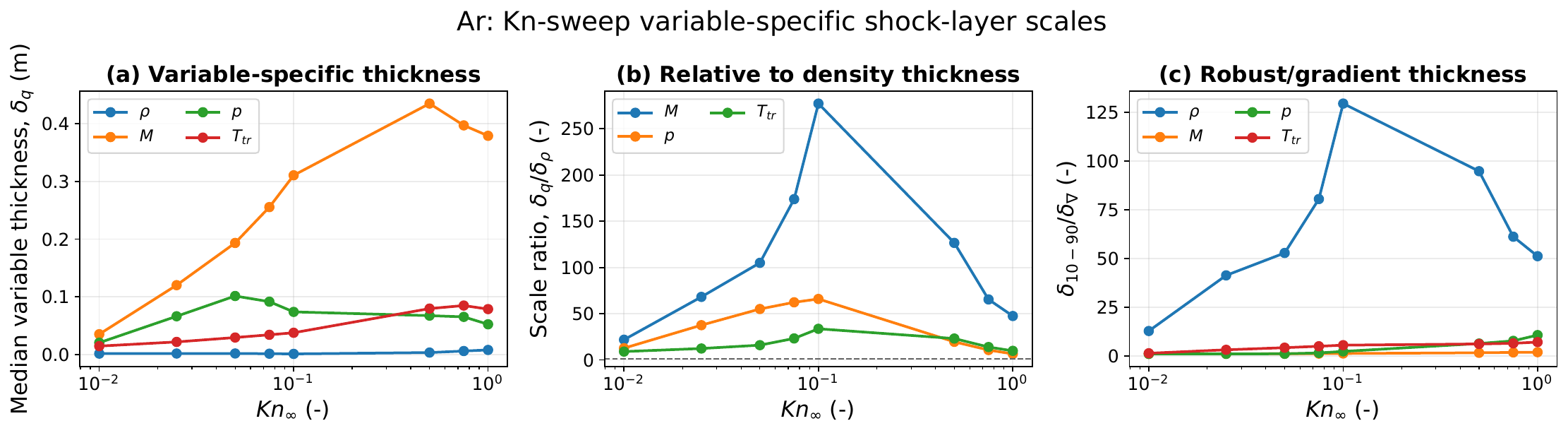}
    \caption{
Variable-specific shock-layer scales for argon in the Knudsen-number sweep.
Panel (a) shows the absolute gradient-based thicknesses for density, Mach
number, pressure and translational temperature. Panel (b) normalizes the same
scales by the density thickness, \(\delta_q/\delta_\rho\), and demonstrates
that density is not a universal reference scale for the full macroscopic
adjustment. Panel (c) compares the robust 10--90 transition thickness with the
local gradient thickness. The large density ratio indicates that a sharp
density-gradient ridge is embedded within a broader finite kinetic transition,
whereas pressure and temperature vary more smoothly. The result shows that
multi-scale shock-layer structure appears even in monatomic argon, before any
rotational or internal-energy relaxation is introduced.
}
    \label{fig:ar_variable_scales_kn}
\end{figure}

The same ray-based extraction can be applied to variables other than density.
Figure~\ref{fig:ar_variable_scales_kn} shows the resulting variable-specific
scales for the argon Knudsen-number sweep. This diagnostic is deliberately
shown for the monatomic gas, because argon has no rotational relaxation mode:
any scale separation observed here is therefore not caused by internal-energy
nonequilibrium, but by the different ways in which density, momentum, pressure
and translational energy respond across a finite kinetic shock layer.

Panel (a) shows that the absolute thicknesses do not grow at the same rate.
The density thickness remains the smallest scale over most of the sweep,
because it is tied to the strongest density-gradient ridge. By contrast, the
Mach-number thickness becomes much larger. This is expected because the Mach
number depends on both the flow speed and the local speed of sound, and hence
on the coupled velocity and translational-temperature relaxation. The velocity
deceleration and thermal adjustment extend over a broader region than the
density-gradient maximum, especially once the mean free path is no longer small
compared with the shock-layer thickness. The pressure and translational
temperature scales occupy an intermediate range. Pressure is not simply a copy
of density: in a rarefied shock layer, \(p\sim \rho T_{tr}\), so density
compression and translational heating can partially compensate or reinforce
each other along different parts of the layer.

Panel (b) makes this scale separation more explicit by normalizing each
variable thickness by \(\delta_\rho\). The large values of
\(\delta_M/\delta_\rho\) and \(\delta_p/\delta_\rho\), particularly in the
intermediate-\(Kn_\infty\) regime, show that density is a poor universal
reference length for the full macroscopic adjustment. Around
\(Kn_\infty\simeq0.1\), the density-gradient layer is still relatively sharp,
whereas the velocity and pressure adjustments have already spread over a much
larger kinetic region. This produces the large peak in the scale ratios. At
higher \(Kn_\infty\), the ratios decrease not because the fields recover a
single-scale structure, but because the density reference scale itself becomes
large and diffuse as collisional localization breaks down. Thus the apparent
reduction of \(\delta_q/\delta_\rho\) at the largest \(Kn_\infty\) should be
interpreted as the broadening of the density layer, not as restoration of
continuum-like similarity.

Panel (c) compares the gradient-based thickness with the robust 10--90
transition thickness. The large values of
\(\delta_{10-90}/\delta_{\nabla}\) for density indicate that the density profile
contains a sharp local ridge embedded in a much broader finite-thickness
transition. In other words, the maximum gradient identifies the most intense
part of the compression layer, but it does not represent the full spatial
extent of the density adjustment. The pressure and temperature ratios are much
smaller because these fields vary more smoothly and are less dominated by a
single narrow ridge. This distinction is important for the interpretation of
high-\(Kn_\infty\) shock metrics: the density-front location and
\(\delta_\rho\) are useful local diagnostics, but they should not be mistaken
for a complete shock-layer thickness when the profile has long kinetic tails.

Overall, figure~\ref{fig:ar_variable_scales_kn} shows that rarefaction creates
a hierarchy of macroscopic length scales even in monatomic argon. The density
compression, velocity/Mach deceleration, pressure adjustment and translational
heating do not share one universal thickness. This supports the central
interpretation of the paper: the inflated bow shock is not simply a translated
and rescaled continuum shock, but a finite kinetic layer in which different
moments of the distribution function relax over different distances.

\begin{figure}
    \centering
    \begin{subfigure}{0.98\textwidth}
        \centering
        \includegraphics[
            width=\textwidth,
            trim={0pt 0pt 0pt 30pt},
            clip
        ]{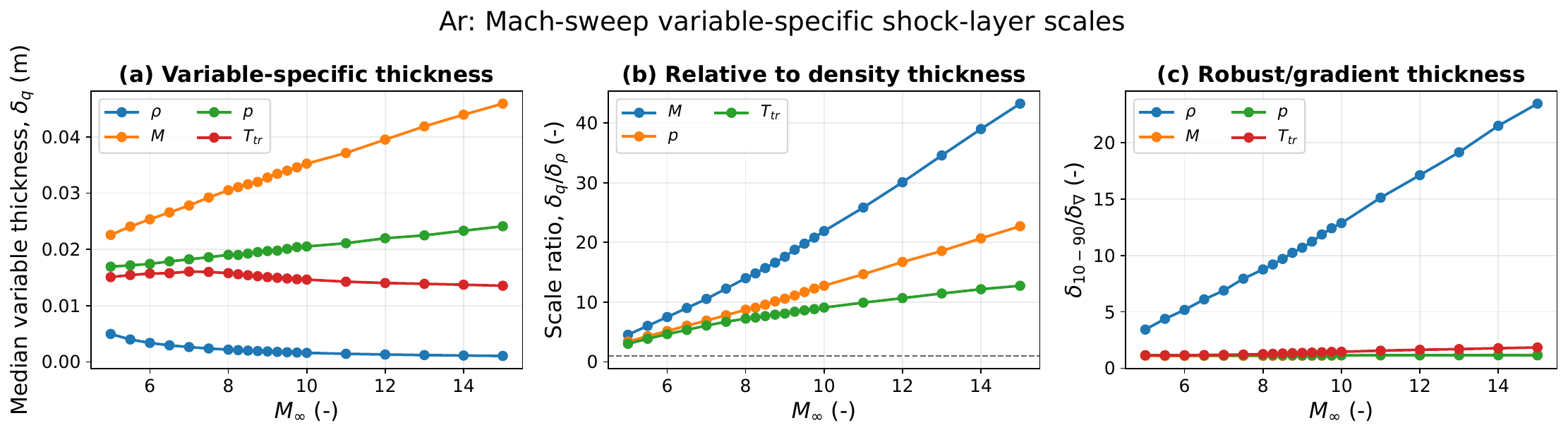}
        \caption{Argon.}
    \end{subfigure}

    \vspace{0.35em}

    \begin{subfigure}{0.98\textwidth}
        \centering
        \includegraphics[
            width=\textwidth,
            trim={0pt 0pt 0pt 30pt},
            clip
        ]{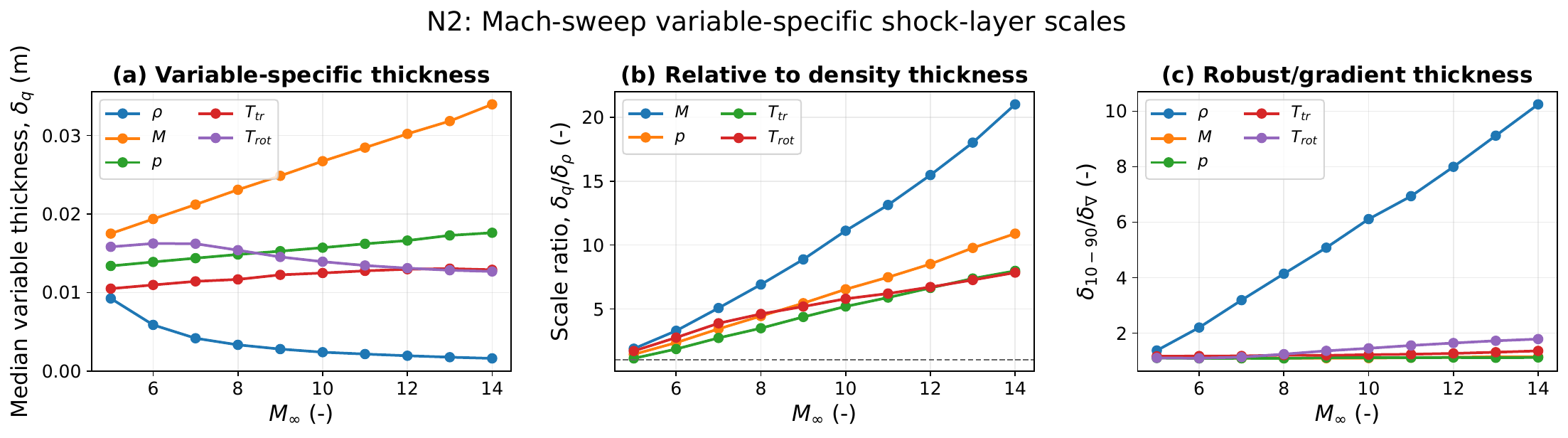}
        \caption{Nitrogen.}
    \end{subfigure}
    \caption{
Variable-specific shock-layer scales in the Mach-number sweep at
\(Kn_\infty=0.01\). Panel (a) shows the absolute gradient-based thicknesses of
density, Mach number, pressure, translational temperature and, for nitrogen,
rotational temperature. Panel (b) normalizes these scales by the density
thickness. The increase of \(\delta_q/\delta_\rho\) with \(M_\infty\) occurs
because the density-gradient layer sharpens as the shock strengthens, whereas
velocity, pressure and thermal relaxation remain distributed over broader
distances. Panel (c) compares the robust 10--90 transition width with the local
gradient thickness. The growing density ratio indicates that a narrow
maximum-gradient ridge is embedded within a broader density transition, while
pressure and temperature vary more smoothly. Thus the Mach sweep preserves a
coherent attached shock layer, but the density scale becomes increasingly
localized relative to the other macroscopic adjustment scales.
}
    \label{fig:mach_variable_scales}
\end{figure}

For completeness, figure~\ref{fig:mach_variable_scales} reports the
variable-specific scales in the Mach-number sweep at fixed
\(Kn_\infty=0.01\). This figure should be interpreted together with
figure~\ref{fig:mach_density_metrics}. At this low rarefaction level, the
density-compression layer remains localized, but increasing \(M_\infty\)
changes how different macroscopic variables adjust across the shock. The
density thickness decreases with Mach number because the stronger shock
produces a sharper density-gradient ridge. This decrease of \(\delta_\rho\)
is the reference against which the other scales are normalized.

The important feature is that the other variables do not decrease in the same
way. The Mach-number thickness increases relative to the density thickness,
especially in nitrogen, because the Mach number depends simultaneously on flow
deceleration and translational heating through the local sound speed. These two
processes need not occur over the same distance as the density jump. Pressure
also develops a larger relative scale than density, since \(p\sim \rho T_{tr}\)
and therefore combines density compression with translational thermal
relaxation. The translational-temperature scale is smoother and varies more
weakly, while the nitrogen rotational-temperature scale remains distinct,
reflecting the finite distance required for rotational energy exchange behind
the density front.

Panel (b) therefore does not imply that the Mach sweep destroys the shock
structure in the same way as the Knudsen-number sweep. Rather, because
\(\delta_\rho\) becomes smaller as the shock strengthens, even moderate
absolute differences between the density, pressure, Mach-number and thermal
thicknesses become amplified when expressed as \(\delta_q/\delta_\rho\). This
is why the scale ratios increase with \(M_\infty\), although the contour plots
and the profile-collapse results still show a coherent attached shock layer.

Panel (c) further separates local-gradient thickness from full transition
width. For both gases, the density \(\delta_{10-90}/\delta_{\nabla}\) ratio
increases with \(M_\infty\), showing that the density profile contains a very
sharp maximum-gradient region embedded within a broader finite transition. The
thermal and pressure ratios remain much closer to unity, indicating smoother
monotonic adjustments without the same degree of local ridge sharpening. Thus
the Mach sweep preserves a compact shock layer, but it also reveals that the
density-gradient scale becomes progressively more localized than the velocity,
pressure and thermal relaxation scales. This behaviour is consistent with the
Hornung-type interpretation of Mach variation as primarily a change in
compression ratio and shock strength, rather than a mean-free-path-driven
delocalization of the entire compression layer.

\subsection{Shock-attached similarity and its breakdown}
\label{sec:collapse}

The preceding metrics show that the density-gradient location and thickness
provide useful local measures of the compression layer, but they do not by
themselves establish whether the entire shock layer is governed by one
similarity scale. We therefore test the stronger hypothesis that the shock
sector can be collapsed by using the density-compression centre and the
density-gradient thickness as the common registration variables. Each field is
sampled along body-normal rays in the upstream shock sector, and the profiles
are registered using
\begin{equation}
    \xi_\rho=\frac{s-s_\rho}{\delta_\rho},
    \label{eq:density_attached_coordinate}
\end{equation}
where \(s\) is the distance along the body-normal ray, \(s_\rho\) is the
maximum-density-gradient location, and \(\delta_\rho\) is the local
density-gradient thickness. Each variable is then normalized by its own local
upstream and downstream values across the same density-defined layer. This
construction deliberately uses the density field to align all variables. It
therefore asks whether velocity, pressure and thermal relaxation are slaved to
the density-compression scale, or whether they retain additional length scales
after the density layer has been aligned. The purpose is not to find the
optimal coordinate for each variable separately, but to test whether the
density-compression layer provides a universal organizing coordinate for the
full shock layer. For the highest-\(Kn_\infty\) cases, this coordinate should
be understood as a compression-layer coordinate based on the strongest
density-gradient marker, not as a coordinate attached to a discontinuous shock
surface.

\begin{figure}
    \centering
    \includegraphics[
        width=0.98\textwidth,
        trim={0pt 0pt 0pt 25pt},
        clip
    ]{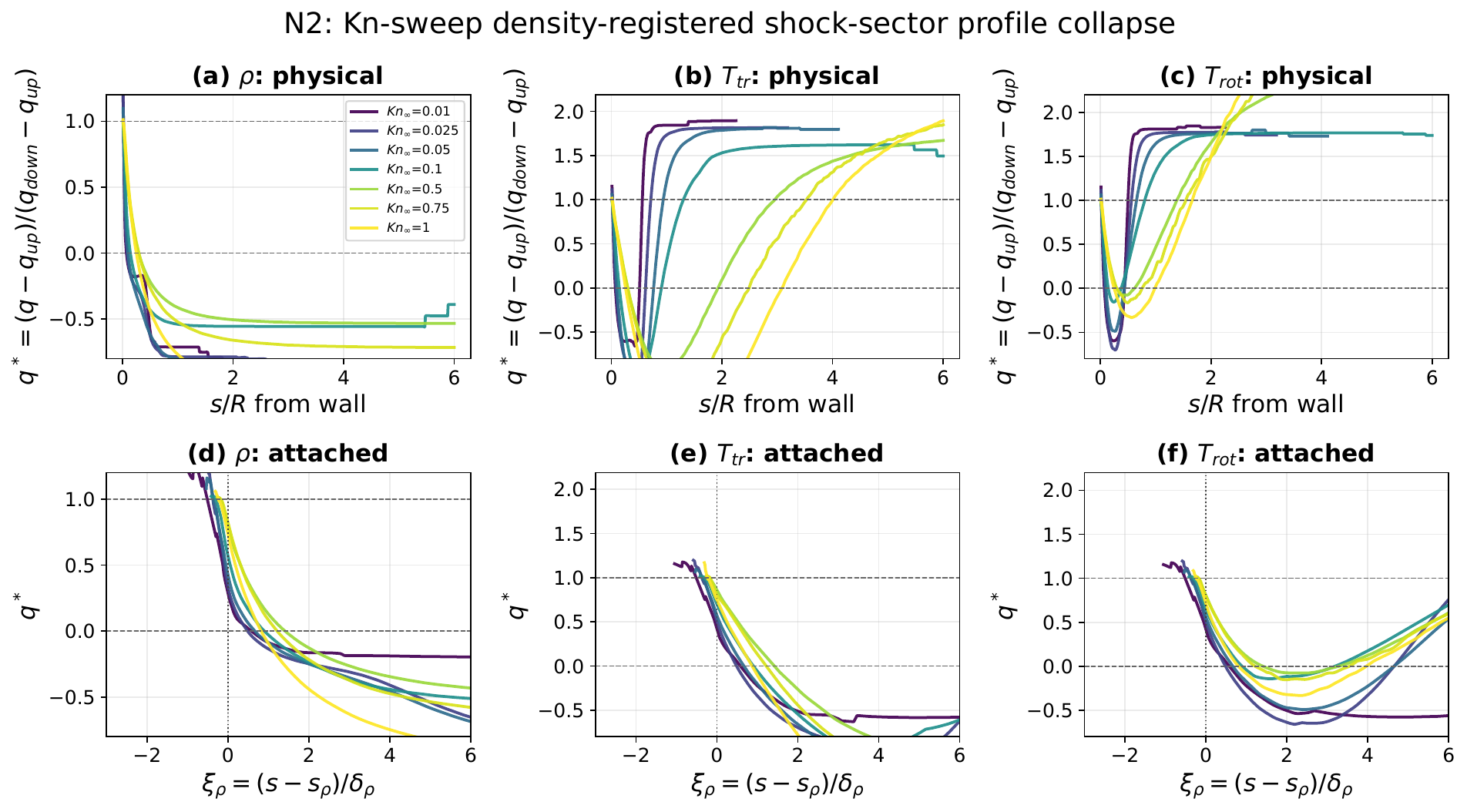}
    \caption{
    Physical and density-registered shock-sector profiles for nitrogen in the
    Knudsen-number sweep at \(M_\infty=10\). The profiles are median profiles
    sampled over body-normal rays in the upstream shock sector. The top row
    shows normalized profiles in the physical coordinate \(s/R\), while the
    bottom row shows the same profiles in the density-attached coordinate
    \(\xi_\rho=(s-s_\rho)/\delta_\rho\). The density field becomes much more
    compact after registration, showing that the geometric inflation of the
    compression layer is largely captured by \(s_\rho\) and \(\delta_\rho\).
    The translational and rotational temperature profiles retain residual
    downstream spread, demonstrating that thermal/internal relaxation is not
    governed by a single density-compression scale.
    }
    \label{fig:n2_kn_profile_collapse}
\end{figure}

Figure~\ref{fig:n2_kn_profile_collapse} shows this test for the nitrogen
Knudsen-number sweep. Nitrogen is the most demanding case because, in addition
to density compression and translational heating, rotational relaxation provides
a separate internal-energy adjustment process. In physical coordinates, the
profiles move and broaden as \(Kn_\infty\) increases, consistent with the
standoff and thickness trends in figure~\ref{fig:kn_density_metrics}. After density-based registration, the main density-transition region is more
closely aligned in the attached coordinate, indicating that a substantial part
of the geometric displacement of the compression layer is represented by
\(s_\rho\) and \(\delta_\rho\). The collapse is not complete: the registered
density profiles still retain downstream tail differences, especially at larger
\(Kn_\infty\). The thermal fields show a similar alignment of the leading
transition near \(\xi_\rho=0\), but their downstream relaxation portions remain
more strongly parameter-dependent. Thus the profile registration removes the
dominant geometric shift, while the residual spread of \(T_{tr}\) and
\(T_{rot}\) indicates additional thermal/internal relaxation scales beyond the
density-compression coordinate.

The failure of the thermal profiles to collapse completely is consistent with
the variable-scale analysis in figures~\ref{fig:ar_variable_scales_kn} and
\ref{fig:mach_variable_scales}. Density marks the most intense compression
ridge, but thermal and internal-energy variables respond to both compression
and post-shock relaxation. As \(Kn_\infty\) increases, the collision length
becomes comparable to the shock-layer scale, so equilibration no longer occurs
within a narrow density-gradient layer. The registered profiles therefore show
the same physics as the thickness metrics: the high-\(Kn_\infty\) bow shock is
not simply a displaced and broadened version of a low-\(Kn_\infty\) shock, but
a finite kinetic compression--relaxation layer with multiple macroscopic
scales.

\begin{figure}
    \centering
    \includegraphics[
        width=0.98\textwidth,
        trim={0pt 0pt 0pt 25pt},
        clip
    ]{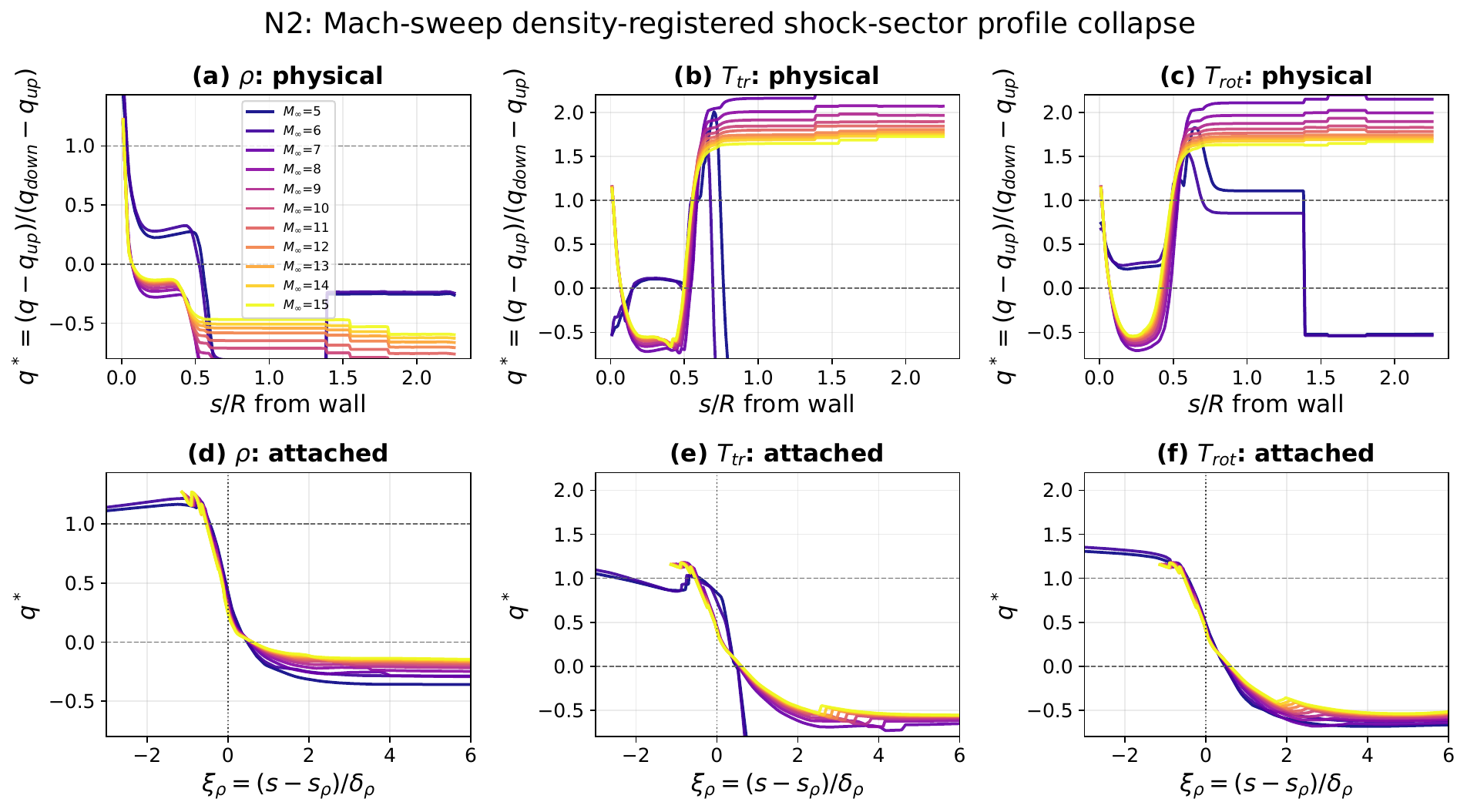}
    \caption{
    Physical and density-registered shock-sector profiles for nitrogen in the
    Mach-number sweep at fixed \(Kn_\infty=0.01\). The top row shows normalized
    profiles in the physical coordinate \(s/R\), and the bottom row shows the
    same profiles in the density-attached coordinate
    \(\xi_\rho=(s-s_\rho)/\delta_\rho\). Compared with the Knudsen-number
    sweep, density-based registration gives a more coherent alignment of the
    density transition, indicating that changing \(M_\infty\) at low rarefaction
    primarily changes shock strength, curvature and thermodynamic level rather
    than producing broad kinetic delocalization. The low-Mach translational
    temperature profiles are more distinct because the translational-temperature
    jump is weaker and the wall-adjacent/post-shock thermal levels have greater
    relative influence in the normalized variable \(q^*\). The thermal panels
    are therefore interpreted as diagnostic evidence of finite relaxation and
    normalization sensitivity, not as evidence of a separate numerical failure
    or of a universal thermal collapse. The low-Mach \(T_{tr}\) curves are more sensitive to the local
upstream/downstream normalization because the thermal jump is weaker and the
thermal-transition location differs more strongly from the high-Mach cases.
Dimensional \(T_{tr}\) profiles remain monotonic with \(M_\infty\); the
normalized low-Mach behaviour is therefore interpreted as a diagnostic
normalization sensitivity of the weak-shock thermal layer.
    }
    \label{fig:n2_mach_profile_collapse}
\end{figure}

The corresponding Mach-number sweep is shown in
figure~\ref{fig:n2_mach_profile_collapse}. This case provides the complementary
limit to figure~\ref{fig:n2_kn_profile_collapse}. Here \(Kn_\infty=0.01\) is
fixed, so the mean free path remains small relative to the body scale and the
shock layer remains comparatively localized. Increasing \(M_\infty\) changes
the compression ratio, shock curvature and standoff distance, as discussed in
figures~\ref{fig:mach_density_sweep} and \ref{fig:mach_density_metrics}, but it
does not introduce the same mean-free-path-driven delocalization seen in the
Knudsen-number sweep. Consequently, density-based registration produces a more
coherent family of density profiles than in the Knudsen-number sweep.

The thermal profiles in figure~\ref{fig:n2_mach_profile_collapse}, especially
\(T_{tr}\) for the lowest Mach numbers, should be interpreted with this
normalization in mind. At \(M_\infty=5\)--6 the shock is weaker, the
translational-temperature jump across the density-defined layer is smaller, and
the imposed diffuse-wall temperature is comparatively more influential relative
to the shock-heated gas than it is at larger Mach numbers. In nitrogen, part of
the translational energy is also transferred into rotation over a finite distance
behind the density front. The normalized quantity
\(q^*=(q-q_{up})/(q_{down}-q_{up})\) therefore amplifies the low-Mach
translational-temperature response: small differences in the downstream thermal
plateau and in the wall-adjacent portion of the ray appear as a larger departure
in the normalized \(T_{tr}\) profiles. This behaviour is not used here as a
separate low-Mach thermal-similarity regime; it indicates that thermal
normalization and relaxation are more sensitive than density compression when
the shock heating is weak.

The residual differences in figure~\ref{fig:n2_mach_profile_collapse} should
therefore be interpreted differently from those in
figure~\ref{fig:n2_kn_profile_collapse}. In the Mach sweep, the spread mainly
reflects changes in shock strength, thermal jump amplitude and post-shock
thermodynamic level at fixed low rarefaction. The density front remains a
meaningful organizing structure, and the profile family is closer to a single
attached-layer description than in the Knudsen-number sweep. In the
Knudsen-number sweep, by contrast, the spread reflects the emergence of a
kinetic relaxation length that is not captured by density registration alone.
The comparison of figures~\ref{fig:n2_kn_profile_collapse} and
\ref{fig:n2_mach_profile_collapse} therefore separates two mechanisms:
Mach-number variation changes the strength of a coherent shock and the relative
thermal jump scale, whereas Knudsen-number variation changes the kinetic nature
of the shock layer itself.

\subsection{Compact multi-modal structure}
\label{sec:pod}

The profile-collapse results show that density registration removes much of
the geometric displacement and broadening of the shock layer, but they also
show that non-density variables retain residual structure. We therefore use
proper orthogonal decomposition (POD) as a parameter-space compactness
diagnostic. The snapshots are indexed by \(Kn_\infty\) or \(M_\infty\), rather
than by time, so the POD modes are not interpreted as temporal instability
modes. Following the standard data-based interpretation of POD, the singular
values identify how many coherent spatial directions are required to represent
the registered family of DSMC fields. In the present usage, ``nearly rank one'' means that the leading mode captures
almost all of the fluctuation variance. For the maximum-density-gradient
registered density fields considered here, this corresponds to \(E_1=0.974\)
for argon and \(E_1=0.986\) for nitrogen in the Knudsen-number sweep
(table~\ref{tab:pod_compactness}), i.e. \(97.4\)--\(98.6\%\) of the variance. Slower modal decay indicates that additional coherent structures
remain after registration. The interpretation is physical only at the level of
the organized leading patterns: density modes primarily describe residual
compression-layer displacement and broadening; Mach-number modes include the
combined effects of velocity deceleration, sound-speed variation and shock
curvature; translational-temperature modes reflect shock heating and finite
wall-temperature interaction; and rotational-temperature modes in nitrogen
measure the additional internal-energy relaxation scale. Higher-order modes are
not interpreted individually because they can contain contributions from DSMC
sampling scatter, finite snapshot number and details of the common-support mask.

The two-dimensional POD is constructed as follows. For each variable and each
case, the registered field \(q^*(\theta,\xi_q)\) is sampled on the common
\(65\times260\) \((\theta,\xi_q)\) grid described above. Points that are outside
the common valid support for more than 30\% of the snapshots are excluded from
the inner product, which prevents extrapolated white regions from influencing
the modes. Each remaining two-dimensional field is flattened into a row vector,
the snapshot mean is subtracted, and singular-value decomposition (SVD) is applied to
the matrix \(\boldsymbol{X}\in\mathbb{R}^{N_s\times N_p}\), where \(N_s\) is the
number of snapshots and \(N_p\) is the number of retained valid grid points. No
additional quadrature weighting is applied; because the attached grid is uniform
in \(\theta\) and \(\xi_q\), the discrete Euclidean inner product is equivalent
to a constant-weight approximation of the attached-coordinate energy. The modal
energy is \(E_i=\sigma_i^2/\sum_j\sigma_j^2\), and the cumulative energy is
\(C_i=\sum_{j=1}^i E_j\), where \(\sigma_i\) is the \(i\)th singular value. These
modal energies are variance fractions in the chosen attached-coordinate
representation; they are not physical kinetic, thermal or thermodynamic energy
norms. A geometry- or Jacobian-weighted inner product would define a different
norm, whereas the present constant-weight norm is used consistently for all
variables so that modal compactness can be compared within the same
registration framework.

The normalization \(q^*=(q-q_{\mathrm{up}})/(q_{\mathrm{down}}-q_{\mathrm{up}})\)
removes the leading jump amplitude and emphasizes the registered transition
shape. The resulting POD energies should therefore be read as compactness
measures of normalized shock-layer structure rather than as amplitudes of
dimensional fluctuations.

The number of parameter snapshots is necessarily limited by the DSMC data set:
the Knudsen-number sweep contains eight snapshots, while the Mach-number sweep
contains eleven snapshots. After common-support masking, the number of retained
spatial grid points changes between variables, but the number of snapshots in
each sweep is fixed. Consequently, the maximum possible number of non-zero POD
modes is small, and
low rank can partly arise from the finite size of the parameter sweep. We
therefore do not interpret the higher modes as a complete statistical basis,
and the 95\% mode count in the tables is only a compactness diagnostic rather
than validation of a predictive reduced-order model. Instead, the POD is used
comparatively: density, pressure, Mach number and thermal variables are
processed with the same grid, masks, normalization and SVD procedure.
Differences in modal decay therefore indicate differences in attached-layer
compactness within the same finite data set.

\appref{app:loo_stability} and \appref{app:loo_reconstruction} report
leave-one-out checks in which each Knudsen-number snapshot is omitted in turn.
These checks are not substitutes for a larger parameter ensemble, but they test
whether the compactness ordering is controlled by one individual snapshot and
whether the retained modes can reconstruct an excluded parameter state. The
results support the central interpretation within the available data: the
density mode is stable to snapshot removal, whereas Mach number and the thermal
variables retain larger leave-one-out variation and therefore should not be
interpreted as single-mode registered families.

\appref{app:modal_sensitivity} also reports two additional checks aimed at
the modal interpretation itself. First, the POD coefficients are plotted as
functions of \(Kn_\infty\), so that the leading modes are connected to coherent
parameter-dependent amplitudes rather than only to singular values. Second, the
POD is repeated with moderate coordinate-space weights and with restricted
attached-coordinate windows. These tests do not change the compactness ordering:
\(\rho\) remains the most compact registered field, \(p\) is intermediate, and
\(M\), \(T_{tr}\) and \(T_{rot}\) retain additional modal content.

Representative block-based intervals for the extracted standoff and thickness
metrics are listed in \appref{app:block_intervals}, and additional
Mach-sweep modal diagnostics are collected in \appref{app:modal}.

\begin{figure}
    \centering
    \begin{subfigure}{0.48\textwidth}
        \centering
        \includegraphics[
            width=\textwidth,
            trim={0pt 0pt 0pt 25pt},
            clip
        ]{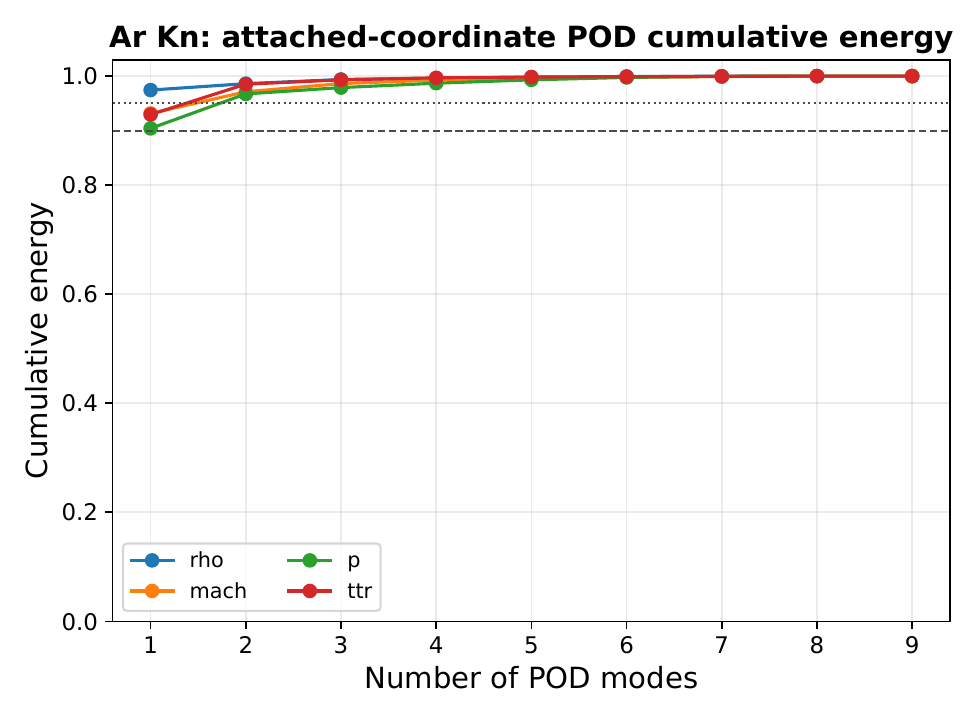}
        \caption{Argon.}
        \label{fig:pod_spectrum_ar_kn}
    \end{subfigure}
    \hfill
    \begin{subfigure}{0.48\textwidth}
        \centering
        \includegraphics[
            width=\textwidth,
            trim={0pt 0pt 0pt 25pt},
            clip
        ]{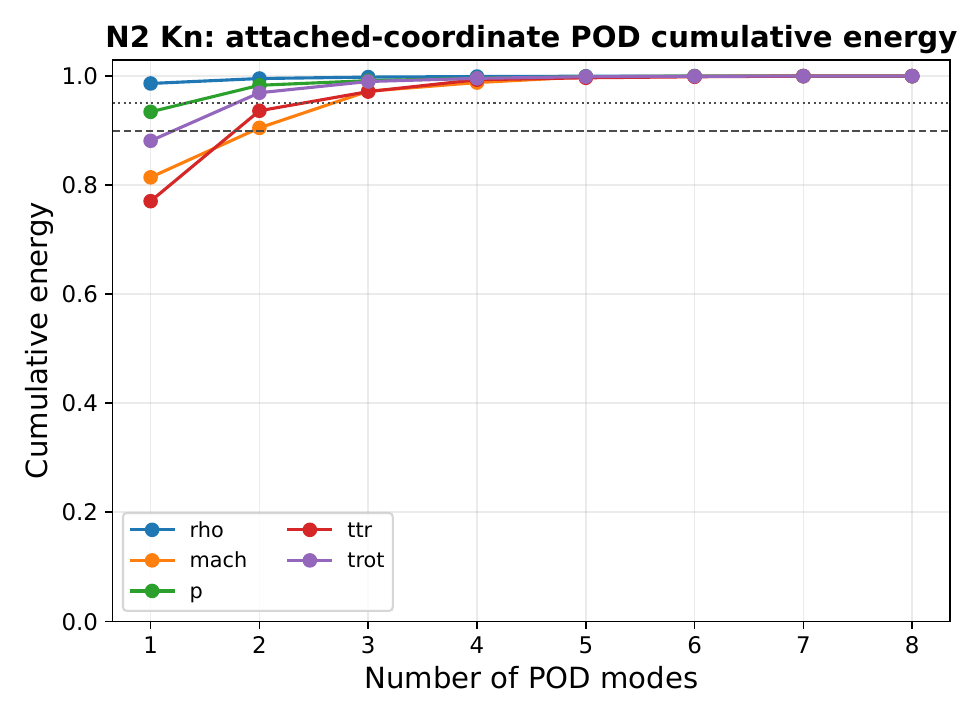}
        \caption{Nitrogen.}
        \label{fig:pod_spectrum_n2_kn}
    \end{subfigure}
    \caption{
    Cumulative POD energy of shock-attached fields in the Knudsen-number
    sweep at \(M_\infty=10\). The density leading mode captures 97.4--98.6\%
    of the registered density variance, while the nitrogen translational and
    rotational temperature fields require additional modes. The spectra are used
    comparatively across variables processed with the same grid, masks and
    normalization, rather than as absolute universal modal energies. They
    quantify the contrast already visible in the density-registered profiles:
    density admits a compact attached-layer description, but
    thermal/internal-energy relaxation retains residual multi-scale structure.
    }
    \label{fig:pod_spectrum_kn_attached}
\end{figure}

Figure~\ref{fig:pod_spectrum_kn_attached} shows the cumulative POD energy for
the two-dimensional shock-attached fields in the Knudsen-number sweep. Within
the present maximum-density-gradient shock-attached registration, the density
field is nearly rank one in this quantitative sense: the
leading mode captures \(97.4\%\) of the argon density variance and
\(98.6\%\) of the nitrogen density variance
(table~\ref{tab:pod_compactness}). This confirms that the density
compression layer admits a compact attached representation. The thermal fields
are less compact. In nitrogen, the leading modes of \(T_{tr}\) and
\(T_{rot}\) capture \(88.7\%\) and \(86.2\%\), respectively, and the
second modes raise the corresponding cumulative energies to \(94.1\%\) and
\(96.5\%\). The first two rotational-temperature modes already exceed the 95\% threshold,
but the need for a second organized mode still shows that the residual
thermal variation is not captured by a single registered template. Hence the
loss of complete similarity is not caused only by shock displacement or
thickness variation; thermal and internal relaxation retain their own modal
content.

\begin{table}
\centering
\caption{POD compactness of shock-attached fields in the Knudsen-number sweep. Here \(E_1\) is the leading fractional POD energy and \(C_2=E_1+E_2\) is the cumulative energy captured by the first two modes.}
\label{tab:pod_compactness}
\begin{tabular}{llccc}
\toprule
Gas & Variable & \(E_1\) & \(C_2\) & Modes for \(95\%\) \\
\midrule
Ar  & \(\rho\)      & 0.974 & 0.986 & 1 \\
Ar  & \(T_{tr}\)   & 0.930 & 0.985 & 2 \\
N2  & \(\rho\)      & 0.986 & 0.996 & 1 \\
N2  & \(M\)         & 0.578 & 0.812 & 4 \\
N2  & \(p\)         & 0.803 & 0.970 & 2 \\
N2  & \(T_{tr}\)   & 0.887 & 0.941 & 3 \\
N2  & \(T_{rot}\)  & 0.862 & 0.965 & 2 \\
\bottomrule
\end{tabular}
\end{table}

Before examining the spatial POD modes, we quantify how much density-attached
registration changes the compactness of the one-dimensional shock-sector
profile families. These profile POD values are not intended to be numerically
identical to the two-dimensional field POD values in table~\ref{tab:pod_compactness},
because the data matrices and normalization domains are different. Table~\ref{tab:n2_registration_gain} compares the leading
POD energy \(E_1\) for nitrogen profiles in the physical coordinate and in the
density-attached coordinate. A table is used rather than a bar chart because
the important information is the selective change in modal compactness for each
variable, not the graphical height of bars. The density profiles become almost
rank one after registration, confirming that the geometric displacement and
broadening of the density-compression layer are largely removed by \(s_\rho\)
and \(\delta_\rho\). Pressure and rotational temperature also gain compactness,
indicating that part of their variation follows the density-layer motion. By
contrast, Mach number and translational temperature do not gain the same
leading-mode compactness. This is consistent with the scale analysis above:
Mach number combines velocity deceleration and local sound-speed variation,
while \(T_{tr}\) reflects translational energy relaxation downstream of the
density ridge. These variables therefore retain structure that is not removed
by density-based registration alone.

These profile POD values are not directly comparable to the two-dimensional
field POD values in table~\ref{tab:pod_compactness}, because the data matrices
and normalization domains are different. The table is included to quantify how
registration changes one-dimensional shock-sector profiles; the two-dimensional
field compactness, including the nitrogen Mach-number and pressure fields shown
in the spectra, is summarized separately in table~\ref{tab:pod_compactness} and
\appref{app:robustness}.

\begin{table}
\centering
\caption{
Effect of density-attached registration on the POD compactness of nitrogen
shock-sector profiles in the Knudsen-number sweep. The columns compare the
leading POD energy in the physical coordinate and in the density-attached
coordinate; \(C_2\) is reported for the density-attached profiles.
}
\label{tab:n2_registration_gain}
\begin{tabular}{lccc}
\toprule
Variable & \(E_1\), physical & \(E_1\), density-attached & \(C_2\), density-attached \\
\midrule
\(\rho\)      & 0.924 & 0.998 & 0.999 \\
\(M\)         & 0.911 & 0.775 & 0.925 \\
\(p\)         & 0.763 & 0.940 & 0.991 \\
\(T_{tr}\)    & 0.885 & 0.853 & 0.981 \\
\(T_{rot}\)   & 0.766 & 0.921 & 0.992 \\
\bottomrule
\end{tabular}
\end{table}

The decrease of the Mach-number leading energy after density-attached
registration, from 0.911 in physical coordinates to 0.775 in attached
coordinates, is a useful counter-example to a purely geometric interpretation of
the registration. Density registration removes the dominant displacement and
broadening of the density-compression layer, but the Mach number is a composite
quantity involving both velocity deceleration and the local speed of sound. Once
the density motion is removed, the remaining Mach-number variability is not
well represented by a single attached template and is redistributed into the
second mode, as reflected by \(C_2=0.925\). This behaviour supports the
conclusion that Mach-number similarity is not completely slaved to the density
front: density alignment removes the dominant geometric motion, but it does not
simultaneously align the velocity deceleration and sound-speed variation that
define the Mach field.

\begin{figure}
    \centering
    \begin{subfigure}{0.48\textwidth}
        \centering
        \includegraphics[
            width=\textwidth,
            trim={0pt 0pt 0pt 25pt},
            clip
        ]{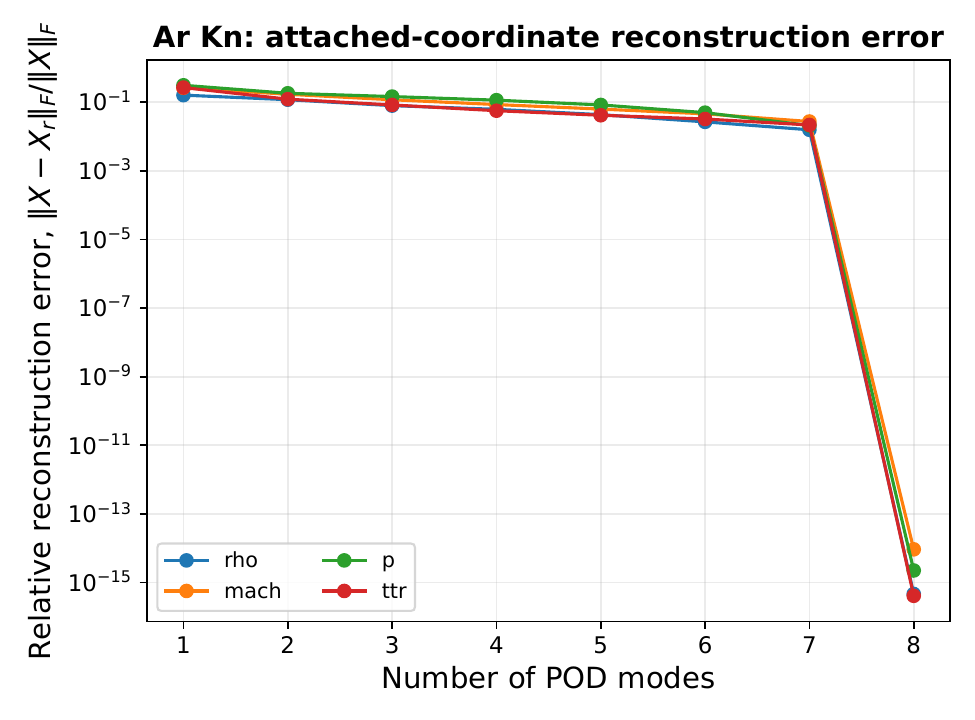}
        \caption{Argon.}
    \end{subfigure}
    \hfill
    \begin{subfigure}{0.48\textwidth}
        \centering
        \includegraphics[
            width=\textwidth,
            trim={0pt 0pt 0pt 25pt},
            clip
        ]{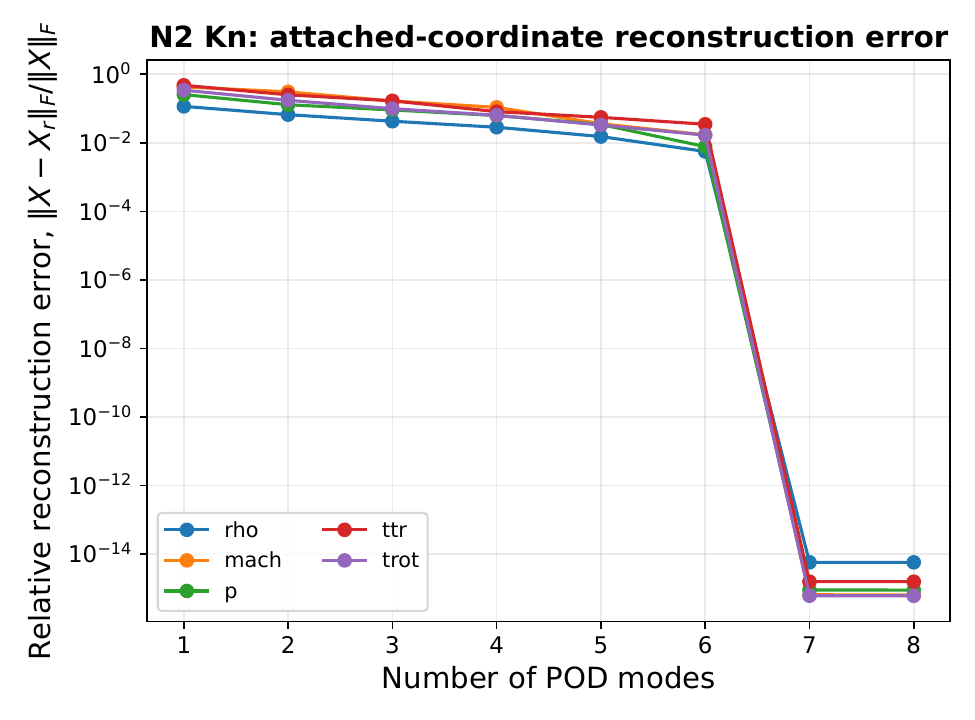}
        \caption{Nitrogen.}
    \end{subfigure}
    \caption{
    Relative POD reconstruction error for shock-attached fields in the
    Knudsen-number sweep. The error curves complement the cumulative-energy
    spectra and show that nitrogen thermal fields retain residual structure
    beyond the leading mode.
    }
    \label{fig:pod_reconstruction_kn_attached}
\end{figure}

Figure~\ref{fig:pod_reconstruction_kn_attached} shows the corresponding
reconstruction error for the two-dimensional fields. The density error
decreases rapidly with the first mode, whereas the nitrogen thermal fields
require additional modes to reach the same accuracy. This reinforces the
conclusion that shock-attached density collapse does not imply full
thermodynamic similarity.

\begin{figure}
    \centering
    \includegraphics[
        width=0.98\textwidth,
        trim={0pt 0pt 0pt 20pt},
        clip
    ]{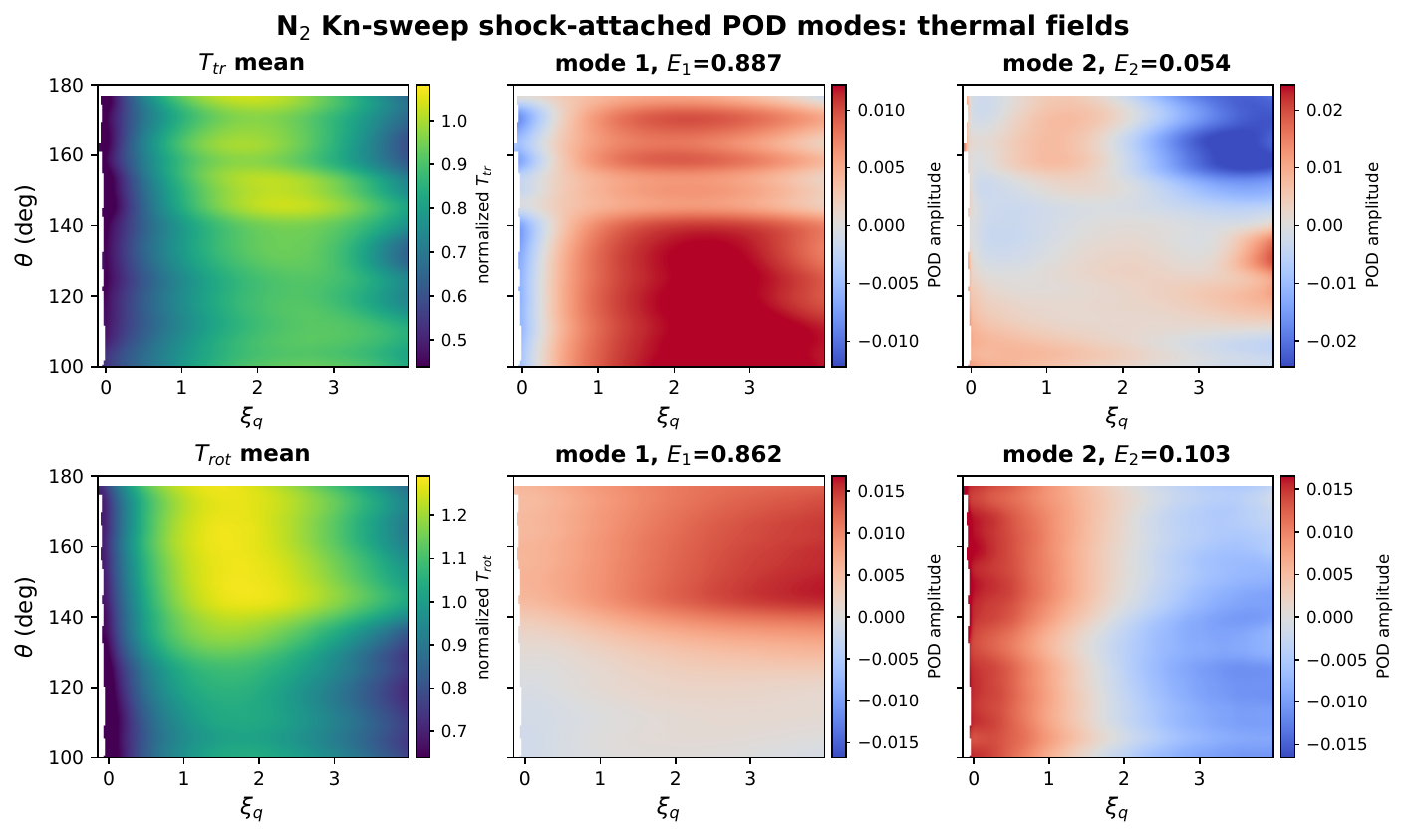}
    \caption{
    Shock-attached POD modes of the nitrogen thermal fields in the
    Knudsen-number sweep. The upper row shows the mean, first mode and second
    mode of the translational-temperature field \(T_{tr}\), while the lower row
    shows the corresponding rotational-temperature structures \(T_{rot}\). The
    mode energies are written on the panels: for \(T_{tr}\), \(E_1=0.887\) and
    \(E_2=0.054\); for \(T_{rot}\), \(E_1=0.862\) and \(E_2=0.103\). White
    regions denote points outside the common valid support of the shock-attached
    interpolation and are not included in the POD inner product. The mean-field
    colour bars show the normalized temperatures, whereas the modal colour bars
    show signed POD amplitudes with symmetric limits. The maps are displayed
    with mild masked smoothing for visualization only; the POD modes and modal
    energies are computed from the unsmoothed registered fields. Because POD
    modes are defined up to an arbitrary sign, the interpretation is based on
    the spatial organization of the patterns rather than on the absolute sign of
    the plotted amplitudes. The modes are spatially organized within the finite
    compression--relaxation layer: the leading mode
    represents the dominant registered relaxation pattern, while the second mode
    captures secondary angular redistribution and downstream-tail variation
    across \(\theta\) and \(\xi_q\). The spatial organization of the leading modes is consistent with coherent
thermal-relaxation structure beyond local DSMC sampling scatter.
    }
    \label{fig:pod_modes_n2_thermal}
\end{figure}

The registration-gain table and reconstruction-error curves show that the
non-density variables retain additional structure after alignment. The spatial
organization of this residual thermal content is shown in
figure~\ref{fig:pod_modes_n2_thermal}. We show only the mean and the first two
thermal POD modes because the spectra and the compactness measures already
establish that the remaining variability is low-dimensional. The important
point is not merely that more than one mode is present, but that the additional
modes are organized within the finite compression--relaxation layer.

The modelling implication is that a reduced-order or neural-operator
representation of rarefied hypersonic bow shocks should not rely only on a
single density-attached coordinate. Such a coordinate is highly effective for
compressing the density field and for removing much of the apparent geometric
inflation, but additional state variables or latent coordinates are needed to
represent velocity, pressure and thermal relaxation with comparable accuracy.
For nitrogen, in particular, the translational and rotational temperatures
retain coherent downstream and angular structures after density registration;
these structures are precisely the components that a single-scale shock-shift
model would miss.

\section{Conclusions}
\label{sec:conclusions}

This study examined whether rarefied hypersonic bow-shock inflation over a
circular cylinder is a simple displacement and broadening of one shock layer,
or whether different moments of the gas distribution retain distinct
attached-layer scales. DSMC data for argon and nitrogen were analysed over a
Knudsen-number sweep at \(M_\infty=10\) and a Mach-number sweep at
\(Kn_\infty=0.01\). At low rarefaction, the ray-based density-gradient ridge
agrees with an independent SWD shock-centre extraction, confirming that a
reproducible shock-front location exists when the shock remains sufficiently
narrow. At higher \(Kn_\infty\), the same diagnostic no longer represents a
unique discontinuity; it identifies the strongest compression location inside a
broad kinetic layer.

The density-based metrics separate the two mechanisms. In the Mach sweep,
\(Kn_\infty\) is fixed and increasing \(M_\infty\) mainly changes the inverse
normal-shock density ratio, shock strength and curvature. The density front
moves closer to the cylinder and the density-gradient layer sharpens while the
shock remains comparatively coherent. In the Knudsen-number sweep,
\(M_\infty\) is fixed, so the inviscid compression parameter is nearly fixed;
the increase in standoff and the high-\(Kn_\infty\) growth of thickness are
therefore kinetic effects associated with the loss of collisional localization.
The intermediate minimum in the effective density thickness reflects the
competition between a decreasing sampled density jump and a weakening maximum
density gradient.

The variable-scale, profile-collapse and POD analyses show that, within the
present two-dimensional cylinder benchmark, gas models and parameter ranges,
the inflated layer is not governed by one universal thickness. Under the
maximum-density-gradient registration used here, the density field becomes
almost rank one, but the profile-level registration test shows that this
compactness is selective: pressure and rotational temperature partly follow the
density-layer motion, whereas Mach number and translational temperature retain
independent modal content. The nitrogen thermal POD modes are spatially
organized within the compression--relaxation layer and are consistent with
coherent relaxation structure beyond local DSMC sampling scatter. Rarefied
bow-shock inflation should therefore be viewed, for the present gases and
parameter ranges, as a coupled compression--relaxation process in which density
collapse and thermal/internal-energy relaxation obey different attached-layer
structures, rather than as a universal single-scale rescaling of a continuum-like
shock.

For modelling, the results suggest that one-coordinate reduced descriptions are
likely to be most effective for density compression. They should not be expected
to transfer unchanged to Mach number or translational/rotational temperature,
for which additional modes, variable-specific relaxation information, or
additional coordinates may be needed.

\section*{Funding}
This research received no specific grant from any funding agency, commercial or
not-for-profit sectors.

\section*{Declaration of interests}
The authors report no conflict of interest.

\section*{Data availability statement}
The DSMC-derived post-processing data, ray-registered modal matrices,
figure-generation scripts and sensitivity-analysis scripts are available upon
reasonable request from the corresponding author.

\section*{Author contributions}
E.R. and A.S.S. contributed equally to the development of the code, simulation
runs, data analysis, and manuscript preparation.

\appendix

\section{Extraction-choice robustness}
\phantomsection
\label{app:robustness}

The main analysis uses one fixed post-processing pipeline so that all variables
are treated consistently. The sensitivity analysis repeats the same extraction
with controlled changes in the angular sector, number of body-normal rays,
attached-grid resolution, display-independent smoothing width and common-support
threshold. This check is designed to separate physically organized modal
differences from changes introduced by reasonable numerical registration
choices within the same maximum-density-gradient definition used in the main
text. In particular, it tracks the leading POD energies of the registered
fields under the same alternatives. Table~\ref{tab:app_robustness_ranges}
reports the resulting ranges for the nitrogen Knudsen-number sweep. The small
spread of the density values and the persistent separation between density and
thermal/internal-energy variables support the comparative interpretation used in
the main text.

\begin{table}
\centering
\caption{Robustness of shock-attached POD compactness to extraction choices for
the nitrogen Knudsen-number sweep. The ranges are taken over the tested
angular-sector, ray-resolution, attached-grid, smoothing and common-support
variants while retaining the maximum-density-gradient marker used in the main
analysis. They are sensitivity ranges for the chosen attached-coordinate
representation, not physical energy uncertainties.}
\label{tab:app_robustness_ranges}
\begin{tabular}{lccc}
\toprule
Variable & Range of \(E_1\) & Range of \(C_2\) & Modes for \(95\%\) \\
\midrule
\(\rho\) & 0.985--0.988 & 0.996--0.997 & 1 \\
\(M\) & 0.563--0.632 & 0.794--0.870 & 4 \\
\(p\) & 0.790--0.827 & 0.966--0.974 & 2 \\
\(T_{tr}\) & 0.865--0.903 & 0.937--0.950 & 2--3 \\
\(T_{rot}\) & 0.813--0.887 & 0.945--0.976 & 2--3 \\
\bottomrule
\end{tabular}
\end{table}

The density field remains nearly rank one for every tested extraction choice
within the main maximum-gradient registration, with \(E_1\) confined to
0.985--0.988 and \(C_2\) to 0.996--0.997. By contrast, the Mach-number and
thermal/internal-energy fields remain less compact. The absolute modal energies
move within the reported ranges, but the ordering of compactness is unchanged:
density is the most compact registered field, pressure is moderately compact,
and Mach number and thermal variables retain additional coherent modal content.
Thus the main conclusion does not rely on a single choice of angular sector,
ray resolution, attached-grid size, smoothing width or common-support threshold.

As a separate check, the density-gradient marker was deliberately replaced by
broader compression-layer markers: a 50\% transition centre with a 10--90
transition width and a gradient-weighted centroid with a gradient-weighted
width. These alternatives do not define the same attached coordinate as the
main analysis and are therefore not used as uncertainties on the main POD
energies. They instead quantify how strongly the modal compactness depends on
using a coordinate tied to the sharpest density-gradient ridge rather than to a
broader transition centre. Table~\ref{tab:app_marker_alternatives} shows that
the numerical energies can change substantially under these alternative
registrations, especially for the density and temperature fields. This is
expected because the half-jump and centroid coordinates align the broad
transition rather than the local compression ridge. Nevertheless, the
alternative-marker test does not convert the Mach-number or thermal fields into
single-mode registered families; these variables continue to require additional
modes. The comparison therefore supports the operational interpretation used
throughout the paper: the maximum-gradient coordinate is a reproducible density
compression-layer coordinate, and the modal energies should be interpreted
within that specified representation.

\begin{table}
\centering
\caption{Effect of deliberately changing the compression-layer marker on the
nitrogen Knudsen-number POD compactness. The first data column gives the range
from table~\ref{tab:app_robustness_ranges} for the main maximum-gradient
registration. The last two columns are alternative-registration diagnostics,
not uncertainty intervals on the main values. Each entry reports \(E_1/C_2\).}
\label{tab:app_marker_alternatives}
\begin{tabular}{lccc}
\toprule
Variable & Max-gradient range & Half-jump marker & Gradient-centroid marker \\
\midrule
\(\rho\)     & 0.985--0.988 / 0.996--0.997 & 0.618 / 0.876 & 0.913 / 0.963 \\
\(M\)        & 0.563--0.632 / 0.794--0.870 & 0.598 / 0.813 & 0.601 / 0.808 \\
\(p\)        & 0.790--0.827 / 0.966--0.974 & 0.572 / 0.870 & 0.807 / 0.927 \\
\(T_{tr}\)   & 0.865--0.903 / 0.937--0.950 & 0.709 / 0.877 & 0.696 / 0.886 \\
\(T_{rot}\)  & 0.813--0.887 / 0.945--0.976 & 0.853 / 0.927 & 0.712 / 0.938 \\
\bottomrule
\end{tabular}
\end{table}

\section{Empirical scaling, uncertainty and finite-snapshot diagnostics}
\phantomsection
\label{app:scaling_support}

This appendix collects the support diagnostics that are first invoked in the
main text after the extraction-choice robustness check. The subsections are
ordered by their first use in the article: empirical scaling of the extracted
metrics, leave-one-out POD stability, leave-one-out reconstruction, modal
coefficient/inner-product/window checks, and finally the representative
block-based intervals available for a limited subset of cases. The empirical
curves are diagnostic summaries of the extracted DSMC trends, not universal
analytical correlations, and the uncertainty intervals are representative
sampling-variability checks rather than a new solver-verification exercise.

\subsection{Empirical diagnostic scaling of the extracted metrics}
\phantomsection
\label{app:empirical_scaling}

\begin{figure}
    \centering
    \includegraphics[
        width=0.98\textwidth,
        trim={0pt 0pt 0pt 0pt},
        clip
    ]{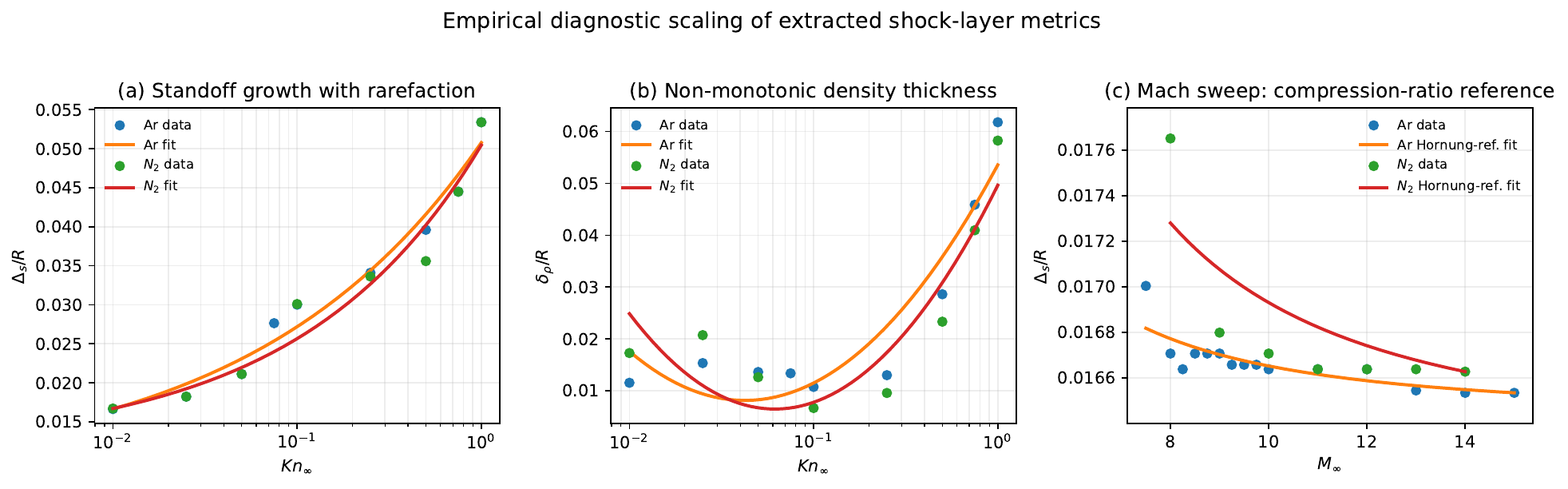}
    \caption{
    Empirical diagnostic scaling of the extracted shock-layer metrics. Panel
    (a) shows the monotonic increase of the density-front standoff with
    \(Kn_\infty\), summarized by anchored power-law fits. Panel (b) shows the
    non-monotonic density-thickness trend, summarized by a quadratic fit in
    \(\log_{10}Kn_\infty\). Panel (c) shows the Mach-sweep standoff against
    \(M_\infty\), with fitted curves based on a Hornung-type compression-ratio
    reference scale. The curves are compact empirical summaries of the DSMC
    trends and are not universal analytical correlations. An automatically
    flagged same-gas Mach-sweep extraction point was excluded from the plotted
    and fitted Mach-sweep data after inspection.
    }
    \label{fig:app_scaling_diagnostic}
\end{figure}

Figure~\ref{fig:app_scaling_diagnostic} summarizes the extracted standoff and
density-thickness metrics using compact empirical fits. The fits are not
proposed as analytical shock laws. Panel (a) shows that the density-front
standoff increases monotonically with \(Kn_\infty\) for both gases, consistent
with a mean-free-path-controlled displacement of the strongest compression
region. Panel (b) summarizes the non-monotonic density-thickness trend: the
intermediate-
\(Kn_\infty\) minimum reflects competition between the decreasing sampled density
jump and the loss of collisional localization. Panel (c) plots the
Mach-sweep standoff against \(M_\infty\), using a Hornung-type
compression-ratio factor only as a low-rarefaction reference scale. One
same-gas Mach-sweep extraction point was automatically flagged as an extreme
outlier and excluded from the displayed and fitted Mach-sweep data; the fit is
therefore used only as a diagnostic summary of the remaining trend.

\subsection{Leave-one-out POD stability}
\phantomsection
\label{app:loo_stability}

The Knudsen-number POD uses a finite set of eight parameter snapshots. To check
whether the compactness ordering is controlled by one individual snapshot,
table~\ref{tab:app_loo_pod_clean} reports leave-one-out ranges obtained by
omitting each snapshot in turn. The density field remains highly stable: the
leading-mode energy stays between 0.983 and 0.987 and the leading-mode shape
similarity remains above 0.995. Pressure also remains comparatively stable and
two-mode compact. Mach number remains less compact for every leave-one-out
subset. The thermal fields show larger variation, particularly \(T_{tr}\),
which is sensitive to removal of the lowest-
\(Kn_\infty\) snapshot. This sensitivity is consistent with the physical
interpretation in the main text: thermal relaxation is not represented by one
universal registered template, and only the organized leading patterns are
interpreted.

\begin{table}
\centering
\caption{Leave-one-out stability of shock-attached POD compactness for the
nitrogen Knudsen-number sweep. Each range is obtained by omitting one
Knudsen-number snapshot at a time. The mode-shape similarity is the absolute
inner product between the baseline and leave-one-out leading modes on their
common support.}
\label{tab:app_loo_pod_clean}
\begin{tabular}{lccc}
\toprule
Variable & Leave-one-out \(E_1\) range & Leave-one-out \(C_2\) range & Leading-mode similarity range \\
\midrule
\(\rho\)     & 0.983--0.987 & 0.993--0.997 & 0.995--1.000 \\
\(M\)        & 0.506--0.686 & 0.786--0.861 & 0.961--0.997 \\
\(p\)        & 0.717--0.830 & 0.958--0.979 & 0.986--1.000 \\
\(T_{tr}\)   & 0.449--0.926 & 0.753--0.967 & 0.485--1.000 \\
\(T_{rot}\)  & 0.589--0.911 & 0.876--0.981 & 0.980--1.000 \\
\bottomrule
\end{tabular}
\end{table}

\subsection{Leave-one-out POD reconstruction}
\phantomsection
\label{app:loo_reconstruction}

\begin{figure}
    \centering
    \includegraphics[
        width=0.72\textwidth,
        trim={0pt 0pt 0pt 0pt},
        clip
    ]{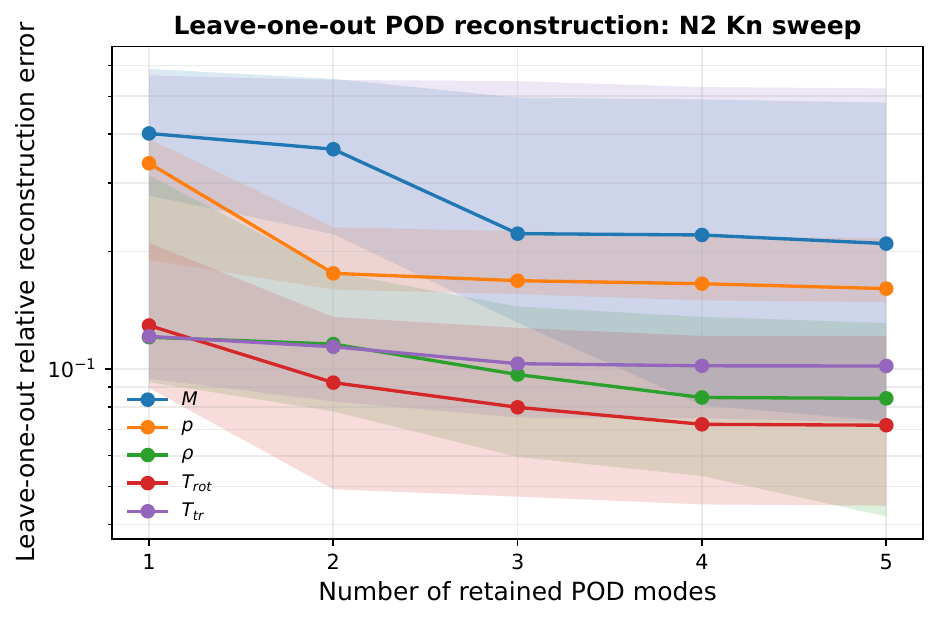}
    \caption{
    Leave-one-out reconstruction error for the nitrogen Knudsen-number sweep.
    Each snapshot is excluded in turn, the POD basis is computed from the
    remaining snapshots, and the excluded field is reconstructed with the first
    \(r\) modes. The solid curves show median relative \(L_2\) error and the
    shaded bands show the 10--90 percentile range across left-out cases. The
    density, pressure and rotational-temperature fields benefit rapidly from
    the first few modes, whereas Mach number and \(T_{tr}\) retain larger
    errors for some excluded states. The result supports a selective, rather
    than universal, low-dimensional attached representation.
    }
    \label{fig:loo_pod_reconstruction}
\end{figure}

The stability ranges above show how the singular values and leading modes
change when one snapshot is removed. A stricter test is to exclude one snapshot,
build the POD basis from the remaining snapshots only, and reconstruct the
excluded field using the first \(r\) modes. Table~\ref{tab:loo_pod_reconstruction}
and figure~\ref{fig:loo_pod_reconstruction} report this leave-one-out
reconstruction test for the nitrogen Knudsen-number sweep. The test should not
be read as a full predictive surrogate validation, because only eight parameter
states are available. It is instead a compact check that the modal compactness
has reconstruction consequences for unseen points within the sampled sweep.

The results are consistent with the selective-compactness interpretation rather
than with a uniformly low-dimensional representation of all variables. Density
has a low one-mode median reconstruction error and improves steadily as more
modes are retained. Mach number is the hardest field to reconstruct, retaining
substantially larger errors even with several modes, which is consistent with
its low \(E_1\) and with the fact that density registration does not align
velocity deceleration and sound-speed variation simultaneously. The pressure and
rotational-temperature fields improve strongly when a second mode is included,
whereas \(T_{tr}\) retains a broad high-error tail for some left-out cases. This
behaviour supports the interpretation that translational thermal relaxation is
not represented by a single density-attached template.

\begin{table}
\centering
\caption{Leave-one-out POD reconstruction errors for the nitrogen
Knudsen-number sweep. For each left-out snapshot, the POD basis is formed from
the remaining snapshots only, and the excluded snapshot is reconstructed with
the first \(r\) modes. Values are median relative \(L_2\) errors over the
left-out cases, with the 10--90 percentile range in parentheses.}
\label{tab:loo_pod_reconstruction}
\begin{tabular}{lccc}
\toprule
Variable & \(r=1\) & \(r=2\) & \(r=3\) \\
\midrule
\(\rho\) & 0.121 (0.0926--0.314) & 0.116 (0.0779--0.177) & 0.0968 (0.0596--0.145) \\
\(M\) & 0.402 (0.278--0.588) & 0.366 (0.222--0.554) & 0.222 (0.132--0.496) \\
\(p\) & 0.337 (0.191--0.389) & 0.176 (0.160--0.231) & 0.168 (0.156--0.225) \\
\(T_{tr}\) & 0.121 (0.0945--0.565) & 0.114 (0.0826--0.551) & 0.103 (0.0752--0.547) \\
\(T_{rot}\) & 0.129 (0.0895--0.211) & 0.0923 (0.0492--0.136) & 0.0798 (0.0471--0.128) \\
\bottomrule
\end{tabular}
\end{table}

\subsection{POD coefficient, weighting and attached-window sensitivity}
\phantomsection
\label{app:modal_sensitivity}

\begin{figure}
    \centering
    \includegraphics[
        width=0.86\textwidth,
        trim={0pt 0pt 0pt 0pt},
        clip
    ]{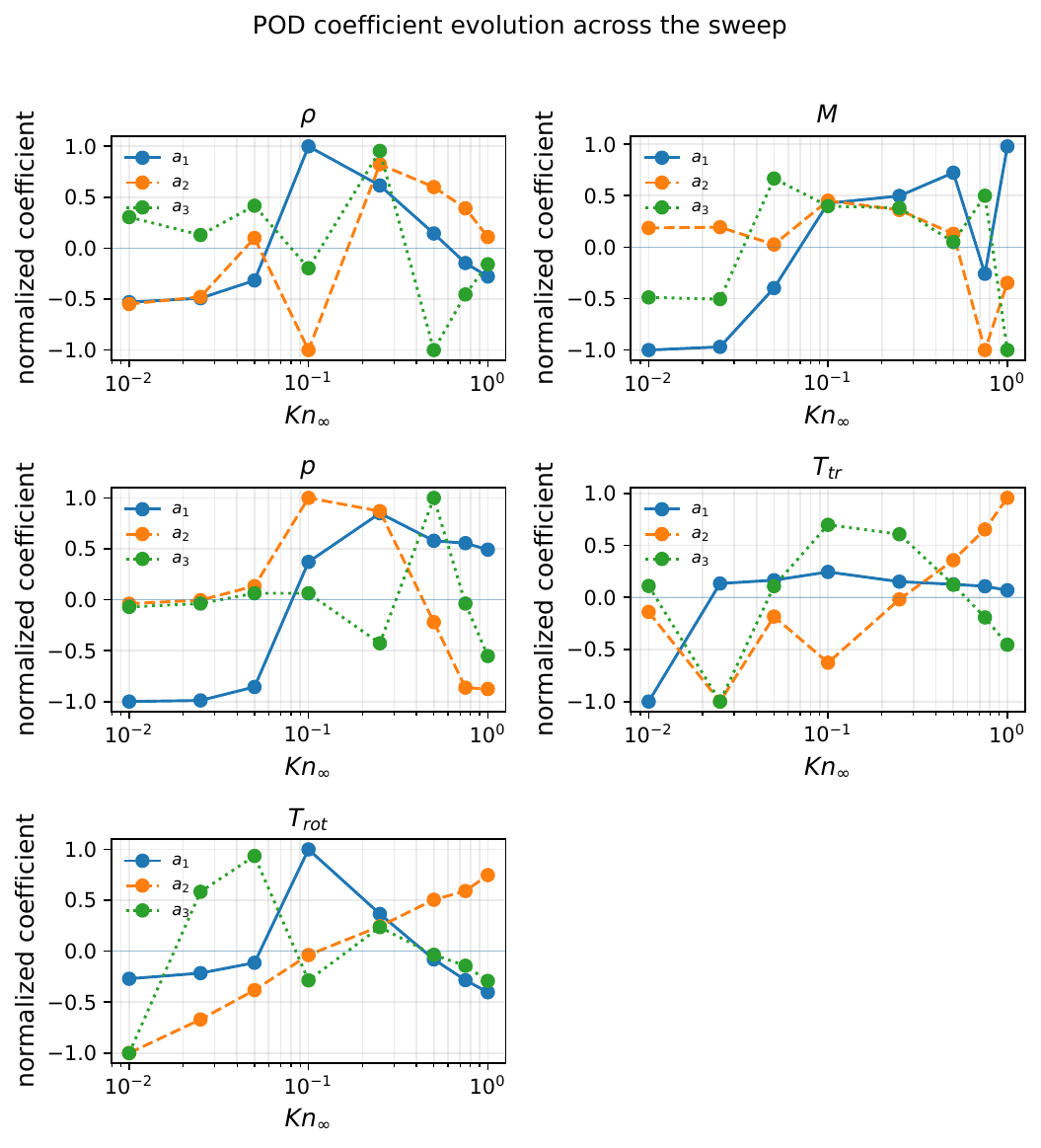}
    \caption{
    Evolution of the first three POD coefficients for the nitrogen
    Knudsen-number sweep. The coefficients are normalized within each variable
    for plotting. The figure is used to show how the leading registered modes
    are activated across parameter space; it is not a temporal modal analysis
    and does not assign dynamical frequencies or growth rates. The multi-mode
    coefficient variation of \(M\), \(T_{tr}\) and \(T_{rot}\) is consistent
    with their slower modal-energy decay in the main text.
    }
    \label{fig:pod_coefficients_vs_parameter}
\end{figure}

The leave-one-out analysis tests whether the finite snapshot set controls the
POD energies. A complementary question is whether the modal interpretation
depends on the particular Euclidean norm or on the far tails of the
attached-coordinate window. Figure~\ref{fig:pod_coefficients_vs_parameter}
therefore reports the first three POD coefficients for the nitrogen
Knudsen-number sweep. The coefficients are normalized separately in each panel
for visual comparison and are plotted only as parameter-space amplitudes, not as
a dynamical modal model. The density and pressure coefficients show that the
leading registered modes are engaged coherently as \(Kn_\infty\) is varied,
whereas the Mach and thermal coefficients require several comparable modal
amplitudes over parts of the sweep. This behaviour is consistent with the
energy spectra: density has a nearly one-mode registered representation, while
Mach number and the thermal/internal-energy fields retain additional organized
directions after density-attached registration.

Table~\ref{tab:app_weight_window_compact} summarizes two further POD
sensitivity checks. In the first, the inner product is repeated with moderate
coordinate-space weights that emphasize the downstream attached coordinate,
the stagnation-side angles or the compression core. In the second, the POD is
recomputed on progressively restricted attached-coordinate windows. These are
not alternative physical energy norms; they are robustness tests for the
compactness ordering. The leading-mode shapes remain very similar to the
baseline modes under all tested weights and windows. The density field remains
nearly rank one, pressure remains two-mode compact, and the Mach-number field
remains the least compact. The translational-temperature field becomes more
compact when the downstream attached-coordinate tail is removed, which indicates
that part of its additional modal content is carried by the extended relaxation
tail. This supports the physical interpretation that thermal relaxation, rather
than only density-front displacement, contributes to the similarity breakdown.

\begin{table}
\centering
\caption{Compact summary of POD sensitivity to coordinate-space weighting and
attached-window restriction for the nitrogen Knudsen-number sweep. The
weighting ranges cover uniform, downstream-\(\xi\), stagnation-side-\(\theta\)
and compression-core weights. The window ranges cover the full
\(-1\leq\xi_q\leq4\) window and three restricted windows. The similarity column
reports the minimum absolute correlation of the leading mode with the
corresponding baseline leading mode.}
\label{tab:app_weight_window_compact}
\begin{tabular}{lcccccccc}
\toprule
& \multicolumn{4}{c}{Weighted inner product} & \multicolumn{4}{c}{Attached-window restriction} \\
\cmidrule(lr){2-5}\cmidrule(lr){6-9}
Variable & \(E_1\) & \(C_2\) & \(N_{95}\) & sim. & \(E_1\) & \(C_2\) & \(N_{95}\) & sim. \\
\midrule
\(\rho\) & 0.984--0.985 & 0.993 & 1 & 1.000 & 0.983--0.984 & 0.993 & 1 & 1.000 \\
\(M\) & 0.539--0.579 & 0.798--0.809 & 4 & 0.994 & 0.566--0.622 & 0.786--0.803 & 4 & 0.992 \\
\(p\) & 0.767--0.801 & 0.964--0.972 & 2 & 0.997 & 0.772--0.832 & 0.966--0.979 & 2 & 0.993 \\
\(T_{tr}\) & 0.873--0.888 & 0.929--0.931 & 3 & 1.000 & 0.878--0.928 & 0.930--0.961 & 2--3 & 1.000 \\
\(T_{rot}\) & 0.831--0.881 & 0.964--0.968 & 2 & 0.998 & 0.850--0.868 & 0.964--0.970 & 2 & 0.991 \\
\bottomrule
\end{tabular}
\end{table}

\subsection{Representative block-based intervals}
\phantomsection
\label{app:block_intervals}

Independent block-averaged fields were available only for a limited subset of
cases. Table~\ref{tab:app_block_ci_clean} reports the corresponding intervals
for the density-front standoff, gradient thickness and 10--90 transition width.
The table is therefore a representative sampling-variability check, not a
complete uncertainty quantification for all simulations. The small spread of
\(\Delta_s/R\) in the listed cases supports the robustness of the extracted
standoff marker in the low-rarefaction regime. The larger spread of the
10--90 width, especially when only two independent blocks are available,
indicates that full-transition widths are more sensitive to the diffuse profile
tails than the local density-gradient marker.

\begin{table}
\centering
\caption{Representative independent-block intervals for density-based
shock-layer metrics. The values are reported as mean \(\pm\) half-width of the
nominal 95\% interval over the available independent field blocks. Only cases
for which such blocks were discoverable in the archived post-processing output
are listed.}
\label{tab:app_block_ci_clean}
\begin{tabular}{llccc}
\toprule
Gas & Case & \(\Delta_s/R\) & \(\delta_\rho/R\) & \(\delta_{10-90}/R\) \\
\midrule
Ar & \(Kn_\infty=0.01,\ M_\infty=10\) & \(0.01665\pm0.00013\) & \(0.0118\pm0.0032\) & \(0.151\pm0.379\) \\
\(N_2\) & \(Kn_\infty=0.01,\ M_\infty=9\) & \(0.01680\pm0.00000\) & \(0.0204\pm0.0000\) & \(0.0892\pm0.0000\) \\
\(N_2\) & \(Kn_\infty=0.01,\ M_\infty=10\) & \(0.01667\pm0.00044\) & \(0.01727\pm0.00004\) & \(0.096\pm0.043\) \\
\bottomrule
\end{tabular}
\end{table}

Because independent time blocks were not available for every case, the main
text avoids assigning statistical confidence intervals to all POD energies. The
post-processing nevertheless uses medians over the body-normal ray sector, and
the robustness checks below vary the extraction choices that most directly
affect the registered fields.

\section{Additional modal diagnostics}
\phantomsection
\label{app:modal}

The main text focuses on the Knudsen-number sweep because rarefaction is the
parameter that destroys the unique shock-front representation. The
Mach-number sweep provides a useful control case: the shock strength and
compression ratio change, but the flow remains at fixed low
\(Kn_\infty=0.01\), where collisional localization is still strong. The
additional modal diagnostics in this appendix therefore check that the
conclusions drawn from the Knudsen-number sweep are not simply a consequence of
using shock-attached coordinates. They also show which variables remain compact
when only the shock strength is varied.

\begin{figure}
    \centering
    \begin{subfigure}{0.48\textwidth}
        \centering
        \includegraphics[
            width=\textwidth,
            trim={0pt 0pt 0pt 25pt},
            clip
        ]{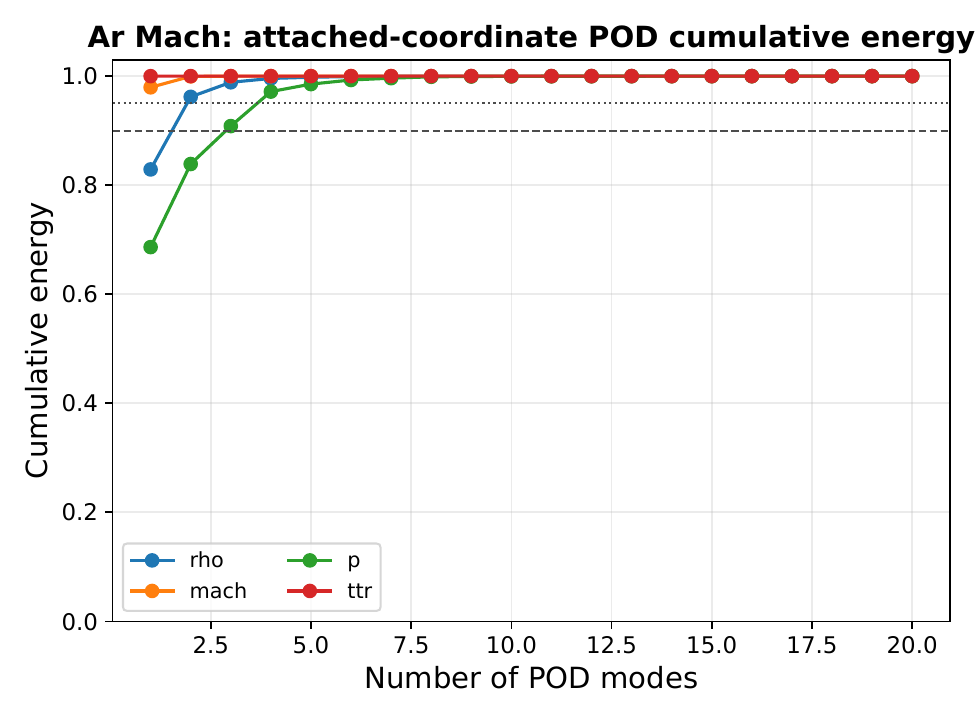}
        \caption{Argon.}
    \end{subfigure}
    \hfill
    \begin{subfigure}{0.48\textwidth}
        \centering
        \includegraphics[
            width=\textwidth,
            trim={0pt 0pt 0pt 25pt},
            clip
        ]{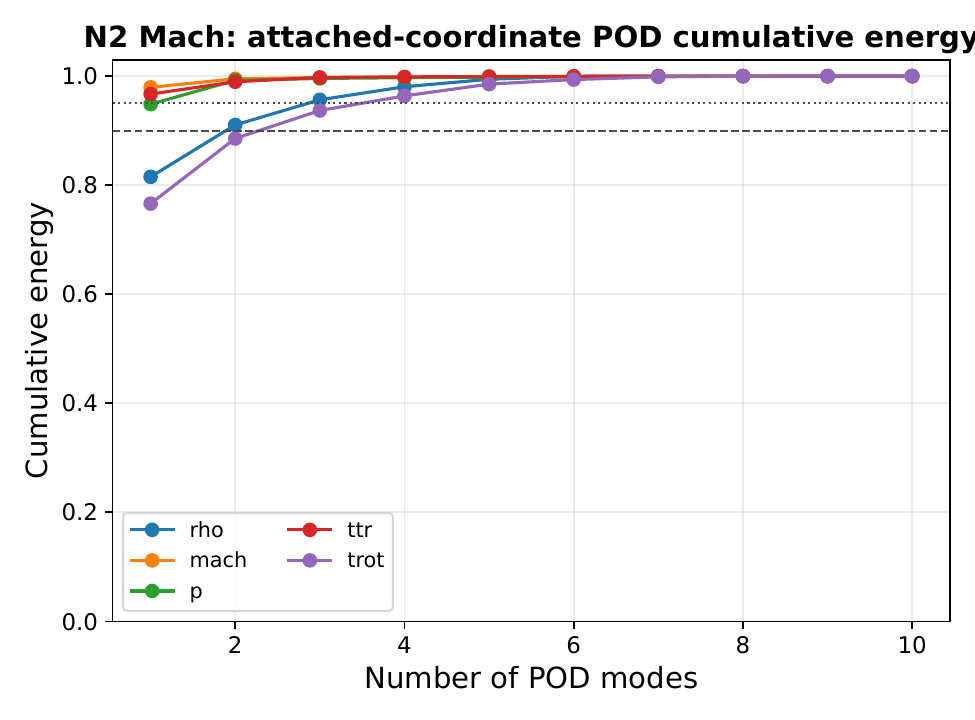}
        \caption{Nitrogen.}
    \end{subfigure}
    \caption{
    Cumulative POD energy of shock-attached fields in the Mach-number sweep at
    fixed \(Kn_\infty=0.01\). The density field remains compact after
    registration, confirming that the Mach sweep preserves a coherent
    compression-layer template. The argon thermal field is also represented by
    a small number of modes, whereas the nitrogen rotational temperature
    requires additional modal content because rotational relaxation introduces
    a thermodynamic scale that is not completely slaved to the density front.
    }
    \label{fig:app_pod_spectrum_mach_attached}
\end{figure}

Figure~\ref{fig:app_pod_spectrum_mach_attached} reports the cumulative POD
energy for the shock-attached Mach-number sweep. The density fields remain
highly compact after registration, consistent with the density-contour and
profile-collapse results in figures~\ref{fig:mach_density_sweep} and
\ref{fig:n2_mach_profile_collapse}. The argon thermal field is also compact,
indicating that, for the monatomic gas, changes in Mach number at low
rarefaction can be represented largely by a small number of attached-layer
structures. Nitrogen is more demanding: the rotational-temperature spectrum
decays more slowly, showing that rotational relaxation still introduces a
secondary thermodynamic scale even when the density front remains coherent.
Thus the Mach sweep is more compact than the Knudsen sweep in the sense of
preserving a localized compression layer, but it is not perfectly single-scale
for every thermodynamic variable.

\begin{figure}
    \centering
    \begin{subfigure}{0.48\textwidth}
        \centering
        \includegraphics[
            width=\textwidth,
            trim={0pt 0pt 0pt 25pt},
            clip
        ]{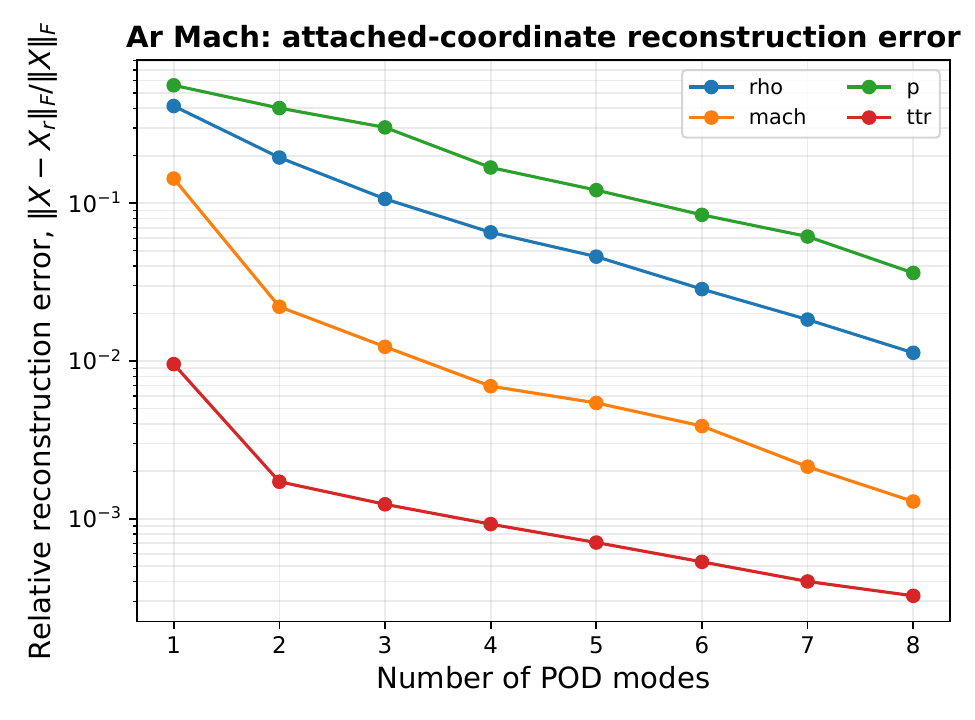}
        \caption{Argon.}
    \end{subfigure}
    \hfill
    \begin{subfigure}{0.48\textwidth}
        \centering
        \includegraphics[
            width=\textwidth,
            trim={0pt 0pt 0pt 25pt},
            clip
        ]{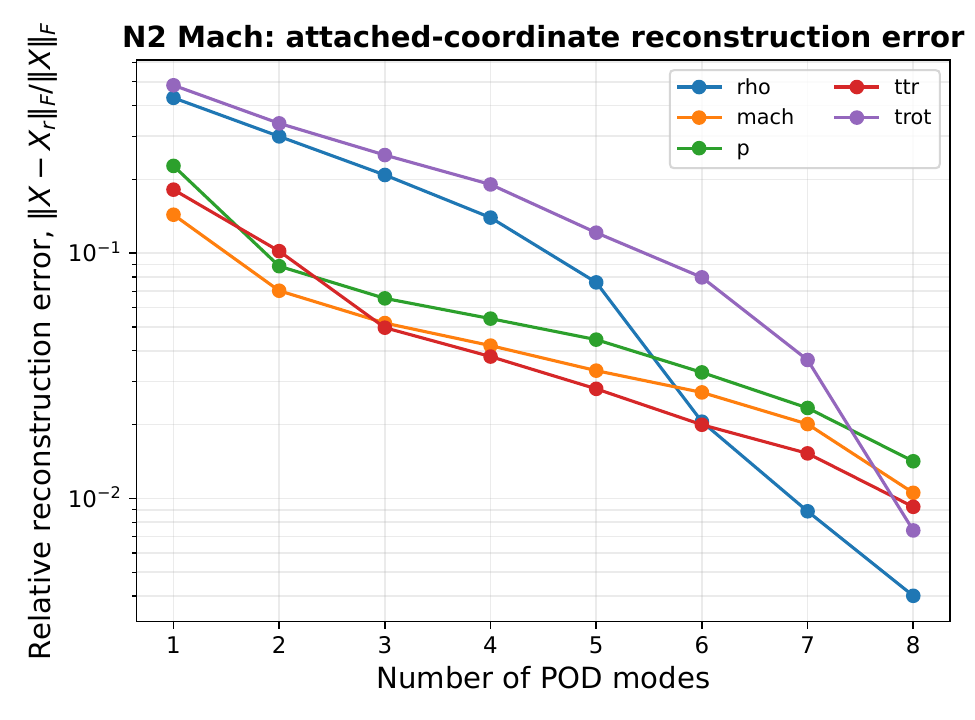}
        \caption{Nitrogen.}
    \end{subfigure}
    \caption{
    Relative reconstruction error for shock-attached fields in the
    Mach-number sweep at \(Kn_\infty=0.01\). The rapid decay of the density
    error confirms that the registered density-compression layer is compact
    when Mach number, rather than Knudsen number, is varied. The slower decay
    of the nitrogen thermal errors indicates residual coherent relaxation
    structure, consistent with the modal spectra in
    figure~\ref{fig:app_pod_spectrum_mach_attached} and with the main-text
    interpretation of thermal/internal-energy similarity breakdown.
    }
    \label{fig:app_pod_reconstruction_mach_attached}
\end{figure}

Figure~\ref{fig:app_pod_reconstruction_mach_attached} gives the corresponding
relative reconstruction errors. These curves are a direct accuracy check on
the energy spectra in figure~\ref{fig:app_pod_spectrum_mach_attached}: a rapid
error decay indicates that a small modal basis reconstructs the registered
Mach-sweep family, while a slower decay identifies residual variable-specific
structure. The density errors fall rapidly for both gases, which reinforces the
interpretation that Mach variation at low rarefaction mainly shifts and
strengthens a coherent density-compression layer. The slower thermal error
decay, especially for nitrogen rotational temperature, is consistent with the
main-text conclusion that internal-energy relaxation can retain coherent
secondary structure even when the density field is compact. The appendix
therefore supports, rather than replaces, the main result: rarefaction is what
produces the broad kinetic delocalization, while gas-dependent thermal
relaxation controls the remaining modal content after registration.

\clearpage

\bibliographystyle{plainnat}
\bibliography{references}

@book{Bird1994,
  author    = {Bird, Graeme A.},
  title     = {Molecular Gas Dynamics and the Direct Simulation of Gas Flows},
  publisher = {Clarendon Press},
  address   = {Oxford},
  year      = {1994}
}

@book{Cercignani2000,
  author    = {Cercignani, Carlo},
  title     = {Rarefied Gas Dynamics: From Basic Concepts to Actual Calculations},
  publisher = {Cambridge University Press},
  year      = {2000}
}

@book{Sone2007,
  author    = {Sone, Yoshio},
  title     = {Molecular Gas Dynamics: Theory, Techniques, and Applications},
  publisher = {Birkh{\"a}user},
  year      = {2007}
}

@book{Karniadakis2005,
  author    = {Karniadakis, George E. and Beskok, Ali and Aluru, Narayana},
  title     = {Microflows and Nanoflows: Fundamentals and Simulation},
  publisher = {Springer},
  year      = {2005}
}

@article{Akhlaghi2017SWD,
  author  = {Akhlaghi, Hassan and Daliri, Abbas and Soltani, Mohammad Reza},
  title   = {Shock-Wave-Detection Technique for High-Speed Rarefied-Gas Flows},
  journal = {AIAA Journal},
  volume  = {55},
  number  = {11},
  pages   = {3747--3756},
  year    = {2017},
  doi     = {10.2514/1.J055819}
}

@article{Akhlaghi2021ShockPolar,
  author  = {Akhlaghi, Hassan and Roohi, Ehsan and Daliri, Abbas and Soltani, Mohammad-Reza},
  title   = {Shock polar investigation in supersonic rarefied gas flows over a circular cylinder},
  journal = {Physics of Fluids},
  volume  = {33},
  number  = {5},
  pages   = {052006},
  year    = {2021},
  doi     = {10.1063/5.0050571}
}

@article{Roohi2026PoF,
  author  = {Roohi, Ehsan and Shoja-Sani, Ahmad and Ebrahimzadeh Azghadi, Fahimeh},
  title   = {Neural networks for rarefied gas dynamics: Relaxation problem, polyatomic shock waves, and hypersonic cylinder flow},
  journal = {Physics of Fluids},
  volume  = {38},
  pages   = {057108},
  year    = {2026},
  doi     = {10.1063/5.0334590}
}

@article{Taira2017Modal,
  author  = {Taira, Kunihiko and Brunton, Steven L. and Dawson, Scott T. M. and Rowley, Clarence W. and Colonius, Tim and McKeon, Beverley J. and Schmidt, Oliver T. and Gordeyev, Stanislav and Theofilis, Vassilios and Ukeiley, Lawrence S.},
  title   = {Modal Analysis of Fluid Flows: An Overview},
  journal = {AIAA Journal},
  volume  = {55},
  number  = {12},
  pages   = {4013--4041},
  year    = {2017},
  doi     = {10.2514/1.J056060}
}

@article{Lumley1967,
  author  = {Lumley, John L.},
  title   = {The structure of inhomogeneous turbulent flows},
  journal = {Atmospheric Turbulence and Radio Wave Propagation},
  pages   = {166--178},
  year    = {1967}
}

@article{Sirovich1987,
  author  = {Sirovich, Lawrence},
  title   = {Turbulence and the dynamics of coherent structures. Part I: Coherent structures},
  journal = {Quarterly of Applied Mathematics},
  volume  = {45},
  number  = {3},
  pages   = {561--571},
  year    = {1987}
}

@article{Berkooz1993,
  author  = {Berkooz, Gal and Holmes, Philip and Lumley, John L.},
  title   = {The proper orthogonal decomposition in the analysis of turbulent flows},
  journal = {Annual Review of Fluid Mechanics},
  volume  = {25},
  pages   = {539--575},
  year    = {1993},
  doi     = {10.1146/annurev.fl.25.010193.002543}
}

@article{Schmid2010,
  author  = {Schmid, Peter J.},
  title   = {Dynamic mode decomposition of numerical and experimental data},
  journal = {Journal of Fluid Mechanics},
  volume  = {656},
  pages   = {5--28},
  year    = {2010},
  doi     = {10.1017/S0022112010001217}
}

@article{Rowley2009,
  author  = {Rowley, Clarence W. and Mezi{\'c}, Igor and Bagheri, Shervin and Schlatter, Philipp and Henningson, Dan S.},
  title   = {Spectral analysis of nonlinear flows},
  journal = {Journal of Fluid Mechanics},
  volume  = {641},
  pages   = {115--127},
  year    = {2009},
  doi     = {10.1017/S0022112009992059}
}

@article{Towne2018,
  author  = {Towne, Aaron and Schmidt, Oliver T. and Colonius, Tim},
  title   = {Spectral proper orthogonal decomposition and its relationship to dynamic mode decomposition and resolvent analysis},
  journal = {Journal of Fluid Mechanics},
  volume  = {847},
  pages   = {821--867},
  year    = {2018},
  doi     = {10.1017/jfm.2018.283}
}

@article{SharipovKalempa2003,
  author  = {Sharipov, Felix and Kalempa, Denize},
  title   = {Gas flow around a longitudinally moving cylinder in the whole range of the Knudsen number},
  journal = {Journal of Vacuum Science \& Technology A},
  volume  = {21},
  number  = {3},
  pages   = {735--745},
  year    = {2003},
  doi     = {10.1116/1.1560710}
}

@article{Sharipov2011,
  author  = {Sharipov, Felix},
  title   = {Data on the velocity slip and temperature jump on a gas-solid interface},
  journal = {Journal of Physical and Chemical Reference Data},
  volume  = {40},
  pages   = {023101},
  year    = {2011},
  doi     = {10.1063/1.3580290}
}

@article{Huang2012UGKS,
  author  = {Huang, Jin and Xu, Kun and Yu, Pu},
  title   = {A unified gas-kinetic scheme for continuum and rarefied flows II: Multi-dimensional cases},
  journal = {Communications in Computational Physics},
  volume  = {12},
  number  = {3},
  pages   = {662--690},
  year    = {2012},
  doi     = {10.4208/cicp.120211.220911s}
}

@article{Zhu2019UGKWP,
  author  = {Zhu, Yajun and Zhong, Chengwen and Xu, Kun},
  title   = {Unified gas-kinetic wave-particle methods. I. Continuum and rarefied gas flow},
  journal = {Journal of Computational Physics},
  volume  = {383},
  pages   = {190--210},
  year    = {2019},
  doi     = {10.1016/j.jcp.2019.01.023}
}

@book{Roohi2025DSMCBook,
  author    = {Roohi, Ehsan and Akhlaghi, Hassan and Stefanov, Stefan},
  title     = {Advances in Direct Simulation Monte Carlo: From Micro-Scale to Rarefied Flow Phenomena},
  publisher = {Springer Nature Singapore},
  address   = {Singapore},
  year      = {2025},
  doi       = {10.1007/978-981-96-8200-3},
  isbn      = {978-981-96-8200-3}
}

@article{Goshayeshi2015SBTTAS,
  author  = {Goshayeshi, Behnam and Roohi, Ehsan and Stefanov, Stefan},
  title   = {{DSMC} simulation of hypersonic flows using an improved {SBT-TAS} technique},
  journal = {Journal of Computational Physics},
  volume  = {303},
  pages   = {28--44},
  year    = {2015},
  doi     = {10.1016/j.jcp.2015.09.044}
}

@article{Lofthouse2008SlipJump,
  author  = {Lofthouse, Andrew J. and Scalabrin, Leonardo C. and Boyd, Iain D.},
  title   = {Velocity Slip and Temperature Jump in Hypersonic Aerothermodynamics},
  journal = {Journal of Thermophysics and Heat Transfer},
  volume  = {22},
  number  = {1},
  pages   = {38--49},
  year    = {2008},
  doi     = {10.2514/1.31280}
}

@article{Schwartzentruber2007Modular,
  author  = {Schwartzentruber, Thomas E. and Scalabrin, Leonardo C. and Boyd, Iain D.},
  title   = {A Modular Particle--Continuum Numerical Method for Hypersonic Non-Equilibrium Gas Flows},
  journal = {Journal of Computational Physics},
  volume  = {225},
  number  = {1},
  pages   = {1159--1174},
  year    = {2007},
  doi     = {10.1016/j.jcp.2007.01.022}
}

@article{Stefanov2000Fluctuations,
  author  = {Stefanov, Stefan K. and Boyd, Iain D. and Cai, Chunpei},
  title   = {Monte Carlo analysis of macroscopic fluctuations in a rarefied hypersonic flow around a cylinder},
  journal = {Physics of Fluids},
  volume  = {12},
  number  = {2},
  pages   = {487--497},
  year    = {2000}
}

@article{Riabov1999SpinningCylinder,
  author  = {Riabov, Vladimir V.},
  title   = {Aerodynamics of a spinning cylinder in rarefied gas flows},
  journal = {Journal of Spacecraft and Rockets},
  volume  = {36},
  number  = {2},
  pages   = {293--298},
  year    = {1999}
}

@article{John2016InverseMagnus,
  author  = {John, B. and Gu, X. J. and Barber, R. W. and Emerson, D. R.},
  title   = {High-speed rarefied flow past a rotating cylinder: The inverse Magnus effect},
  journal = {AIAA Journal},
  volume  = {54},
  number  = {2},
  pages   = {521--532},
  year    = {2016}
}

@article{John2018SpinningCylinder,
  author  = {John, B. and Gu, X. J. and Emerson, D. R.},
  title   = {Computation of aerodynamic forces under nonequilibrium conditions: Flow past a spinning cylinder},
  journal = {AIAA Journal},
  volume  = {56},
  number  = {1},
  pages   = {198--209},
  year    = {2018}
}

@article{John2019Cylinder,
  author  = {John, B. and Gu, X. J. and Barber, R. W. and Emerson, D. R.},
  title   = {Non-equilibrium effects on flow past a circular cylinder in the slip and early transition regime},
  journal = {Journal of Fluid Mechanics},
  volume  = {871},
  pages   = {626--662},
  year    = {2019}
}

@misc{Klothakis2021KineticFlatPlate,
  author        = {Klothakis, Angelos and Quintanilha, Helio, Jr. and Sawant, Saurabh S. and Protopapadakis, Eftychios and Theofilis, Vassilis and Levin, Deborah A.},
  title         = {Linear stability analysis of hypersonic boundary layers computed by a kinetic approach: A semi-infinite flat plate at {Mach} 4.5 and 9},
  year          = {2021},
  eprint        = {2104.12743},
  archivePrefix = {arXiv},
  primaryClass  = {physics.flu-dyn}
}

@misc{Senkardesler2025MolecularSecondMode,
  author        = {Senkardesler, Mert and Karpuzcu, Irmak T. and Levin, Deborah A. and Theofilis, Vassilis},
  title         = {A Molecular Gas Dynamics Study of Hypersonic Boundary Layer Second {Mack} Mode Instabilities},
  year          = {2025},
  eprint        = {2512.11390},
  archivePrefix = {arXiv},
  primaryClass  = {physics.flu-dyn}
}

@article{Roohi2026NeuralCollisionOperators,
  author  = {Roohi, Ehsan and Shoja-Sani, Ahmad and Stefanov, Stefan},
  title   = {Physics constrained neural collision operators for hard sphere surrogates and ab initio angle prediction in direct simulation Monte Carlo},
  journal = {Physics of Fluids},
  volume  = {38},
  pages   = {000000},
  year    = {2026},
  doi     = {10.1063/5.0328463}
}

@article{Hornung1972Nonequilibrium,
  author  = {Hornung, H. G.},
  title   = {Non-equilibrium dissociating nitrogen flow over spheres and circular cylinders},
  journal = {Journal of Fluid Mechanics},
  volume  = {53},
  number  = {1},
  pages   = {149--176},
  year    = {1972},
  doi     = {10.1017/S0022112072000084}
}

@article{Hornung2019SphereCone,
  author  = {Hornung, H. G. and Schramm, Jan Martinez and Hannemann, Klaus},
  title   = {Hypersonic flow over spherically blunted cone capsules for atmospheric entry. Part 1. The sharp cone and the sphere},
  journal = {Journal of Fluid Mechanics},
  volume  = {871},
  pages   = {1097--1116},
  year    = {2019},
  doi     = {10.1017/jfm.2019.342}
}

@article{Hornung2021WedgeCylinder,
  author  = {Hornung, H. G.},
  title   = {Shock detachment and drag in hypersonic flow over wedges and circular cylinders},
  journal = {Journal of Fluid Mechanics},
  volume  = {915},
  pages   = {A100},
  year    = {2021},
  doi     = {10.1017/jfm.2021.187}
}

@article{Lofthouse2007ContinuumBreakdown,
  author  = {Lofthouse, Andrew J. and Boyd, Iain D. and Wright, Michael J.},
  title   = {Effects of continuum breakdown on hypersonic aerothermodynamics},
  journal = {Physics of Fluids},
  volume  = {19},
  number  = {2},
  pages   = {027105},
  year    = {2007},
  doi     = {10.1063/1.2710289}
}

@misc{Sawant2021KineticDoubleWedge,
  author        = {Sawant, Saurabh S. and Theofilis, Vassilios and Levin, Deborah A.},
  title         = {Kinetic modelling of three-dimensional shock/laminar separation bubble instabilities in hypersonic flows over a double wedge},
  year          = {2021},
  eprint        = {2101.08957},
  archivePrefix = {arXiv},
  primaryClass  = {physics.flu-dyn}
}

@misc{RoohiMahdavi2026Nozzle,
  author        = {Roohi, Ehsan and Mahdavi, Amirmehran},
  title         = {Shock-Centered Low-Rank Structure and Neural-Operator Representation of Rarefied Micro-Nozzle Flows},
  year          = {2026},
  eprint        = {2605.12723},
  archivePrefix = {arXiv},
  primaryClass  = {physics.flu-dyn},
  doi           = {10.48550/arXiv.2605.12723}
}

@article{Jiang2024ShockShock,
    author  = {Jiang, Yazhong and Sun, Xuxu and Niu, Jie and Zhang, Jun},
  title   = {Study of shock-shock interactions in rarefied flows using direct simulation {Monte Carlo} method},
  journal = {Aerospace Science and Technology},
  volume  = {173},
  pages   = {111768},
  year    = {2026},
  doi     = {10.1016/j.ast.2026.111768}
}

@article{Liu2019UGKWP,
  author  = {Liu, Chang and Zhu, Yajun and Xu, Kun},
  title   = {Unified gas-kinetic wave-particle methods I: Continuum and rarefied gas flow},
  journal = {Journal of Computational Physics},
  volume  = {401},
  pages   = {108977},
  year    = {2020},
  doi     = {10.1016/j.jcp.2019.108977}
}

@article{Goshayeshi2015SBT,
  author  = {Goshayeshi, Behnam and Roohi, Ehsan and Stefanov, Stefan},
  title   = {A novel simplified {Bernoulli} trials collision scheme in the direct simulation {Monte Carlo} with intelligence over particle distances},
  journal = {Physics of Fluids},
  volume  = {27},
  number  = {10},
  pages   = {107104},
  year    = {2015},
  doi     = {10.1063/1.4934588}
}

@article{Karpuzcu2025RampSPOD,
  author  = {Karpuzcu, Irmak T. and Senkardesler, Mert and Levin, Deborah A.},
  title   = {On flow unsteadiness in strongly separated high-speed ramp flows using kinetic and data-driven methods},
  journal = {Physics of Fluids},
  volume  = {37},
  number  = {9},
  pages   = {096136},
  year    = {2025},
  doi     = {10.1063/5.0281770}
}

\end{document}